\documentclass[aps,prb,superscriptaddress,twocolumn,floatfix,showpacs]{revtex4}
\usepackage[dvips]{color}
\usepackage{graphicx}
\usepackage{amsmath,amssymb,bm}
\usepackage[hypertex,breaklinks=true,colorlinks=false]{hyperref}

\newcommand{\Svec}{\bm{S}}
\newcommand{\Mvec}{\bm{M}}
\newcommand{\Nvec}{\bm{N}}

\newcommand{\bs}{\mathrm{bs}}

\newcommand{\tw}{\mathrm{tw}}
\newcommand{\dtw}{\mathrm{dtw}}

\newcommand{\thetat}{\tilde{\theta}}

\newcommand{\ua}{\uparrow}
\newcommand{\da}{\downarrow}

\newcommand{\Ocal}{{\cal O}}

\newcommand{\ket}[1]{|\!#1 \,\rangle}

\newcommand{\bol}[1]{\boldsymbol #1}
\newcommand{\up}{\uparrow}
\newcommand{\down}{\downarrow}

\newcommand{\sqp}{{\sqrt{\pi}}}
\newcommand{\sqtp}{{\sqrt{2\pi}}}
\newcommand{\sqfp}{{\sqrt{4\pi}}}

\def\rnum#1{\expandafter{%
\romannumeral #1}}
\def\Rnum#1{\uppercase\expandafter{%
\romannumeral #1}}

\begin{document}

%%%%%%%%%%%%%%%%%%%%%%%%%%%%%%%%%%%%%%%%%%%%%%%%%
% Paper Information
%%%%%%%%%%%%%%%%%%%%%%%%%%%%%%%%%%%%%%%%%%%%%%%%%
\title{
Ground-state phase diagram of a spin-$\frac12$ frustrated ferromagnetic 
XXZ chain: \\
Haldane dimer phase and gapped/gapless chiral phases
}
\author{Shunsuke Furukawa}
\affiliation{Department of Physics, University of Tokyo, 7-3-1 Hongo, Bunkyo-ku, Tokyo 113-0033, Japan}
\author{Masahiro Sato}
\affiliation{Department of Physics and Mathematics, Aoyama Gakuin University, Sagamihara, Kanagawa 252-5258, Japan}
\author{Shigeki Onoda}
\affiliation{Condensed Matter Theory Laboratory, RIKEN, Wako, Saitama 351-0198, Japan}
\author{Akira Furusaki}
\affiliation{Condensed Matter Theory Laboratory, RIKEN, Wako, Saitama 351-0198, Japan}
\date{\today}

\pacs{75.10.Jm, 75.10.Pq, 75.80.+q}
% 75.10.Jm  Quantized spin models
% 75.10.Pq  Spin chain models
% 75.80.+q 	Magnetomechanical and magnetoelectric effects, magnetostriction
%-- 
% 75.40.Cx  Static properties (order parameter, static susceptibility, heat capacities, critical exponents, etc.) 
% 75.40.Gb  Dynamic properties (dynamic susceptibility, spin waves, spin diffusion, dynamic scaling, etc.) 
% 75.40.Mg  Numerical simulation studies
% 77.80.-e 	Ferroelectricity and antiferroelectricity
% 77.22.Ej 	Polarization and depolarization
%--------------------------------------

%%%%%%%%%%%%%%%%%%%%%%%%%%%%%%%%%%%%%%%%%%%%%%%%%
% Abstract
%%%%%%%%%%%%%%%%%%%%%%%%%%%%%%%%%%%%%%%%%%%%%%%%%
\begin{abstract}
The ground-state phase diagram of a spin-$\frac12$ XXZ chain with
competing ferromagnetic nearest-neighbor ($J_1<0$) and
antiferromagnetic second-neighbor ($J_2>0$) exchange couplings is
studied by means of the infinite time evolving block decimation
algorithm and effective field theories.
For the SU(2)-symmetric (Heisenberg) case,
we show that the nonmagnetic phase
in the range $-4<J_1/J_2<0$ has a small but
finite ferromagnetic dimer order.
We argue that this spontaneous dimer order is associated with
effective spin-$1$ degrees of freedom on dimerized bonds, which
collectively form a valence bond solid state as in
the spin-$1$ antiferromagnetic Heisenberg chain 
(the Haldane spin chain).
We thus call this phase the Haldane dimer phase.
With easy-plane anisotropy, the model exhibits a variety of phases
including the vector chiral phase with gapless excitations and
the even-parity dimer and N\'eel phases with gapped excitations,
in addition to the Haldane dimer phase.
Furthermore, we show the existence of gapped phases
with coexisting orders in narrow regions that intervene 
between the gapless chiral phase and any one of Haldane dimer,
even-parity dimer, and N\'eel phases.
Possible implications for quasi-one-dimensional edge-sharing cuprates
are discussed.
\end{abstract}

\maketitle

%%%%%%%%%%%%%%%%%%%%%%%%%%%%%%%%%%%%%%%%%%%%%%%%%
\section{Introduction}
%%%%%%%%%%%%%%%%%%%%%%%%%%%%%%%%%%%%%%%%%%%%%%%%%

%--------------------------------------
% Introduction - frustration, 1D, J1-J2model

The search for novel quantum states in frustrated magnets 
has been a subject of intensive theoretical and experimental research. 
One-dimensional (1D) systems offer unique laboratories for this search, 
as strong fluctuations enhance the tendency toward unconventional
quantum states.\cite{Lecheminant05}  
Among them, the 1D XXZ model with competing nearest-neighbor $J_1$ and
second-neighbor $J_2$ interactions,
defined by the Hamiltonian
\begin{equation}\label{eq:H}
  H={\sum_{n=1}^2\sum_\ell } J_n
  \left( 
     S^x_\ell S^x_{\ell+n} + S^y_{\ell} S^y_{\ell+n} 
   + \Delta S^z_\ell S^z_{\ell+n}
  \right),
\end{equation}
provides a paradigmatic example expected to host rich variety of physics. 
Here ${\bol S}_\ell=(S^x_\ell, S^y_\ell, S^z_\ell)$ represents
the spin-$\frac12$ operator at the site $\ell\in \mathbb{Z}$ 
and $\Delta$ parametrizes the XXZ exchange anisotropy. 
The model has frustration as far as $J_2$ is antiferromagnetic,
irrespective of the sign of $J_1$. 

%--------------------------------------
% Growing interest in the model

Early theoretical studies on the model \eqref{eq:H} mostly considered the case 
when both $J_1$ and $J_2$ are antiferromagnetic.\cite{Majumdar69,Haldane82,Nomura94,White96,Nersesyan98,Hikihara01}  
However, interest is now growing in the case of
ferromagnetic $J_1<0$ and antiferromagnetic $J_2>0$
because of its relevance to quasi-1D edge-sharing cuprates. 
Among such cuprates, LiCu$_2$O$_2$ (Refs.~\onlinecite{Masuda05, Park07}),
LiCuVO$_4$ (Refs.~\onlinecite{Enderle05,Naito07}), 
and PbCuSO$_4$(OH)$_2$ (Ref.~\onlinecite{Yasui11}), for example,  
exhibit multiferroic behaviors,\cite{Tokura10,Cheong07}
i.e.,
spiral magnetic orders and concomitant ferroelectric polarization
at low temperatures.
The negative sign of $J_1$ indeed plays a key role 
in stabilizing the vector chiral order responsible
for these phenomena.\cite{FSO10}  
By contrast, Rb$_2$Cu$_2$Mo$_3$O$_{12}$ (Ref.~\onlinecite{Hase04}) 
shows no sign of magnetic order down to very low temperatures
and may be considered as a candidate system
for a spin liquid or a valence bond solid. 

%============================
\begin{figure}
\begin{center}
\includegraphics[width=0.5\textwidth]{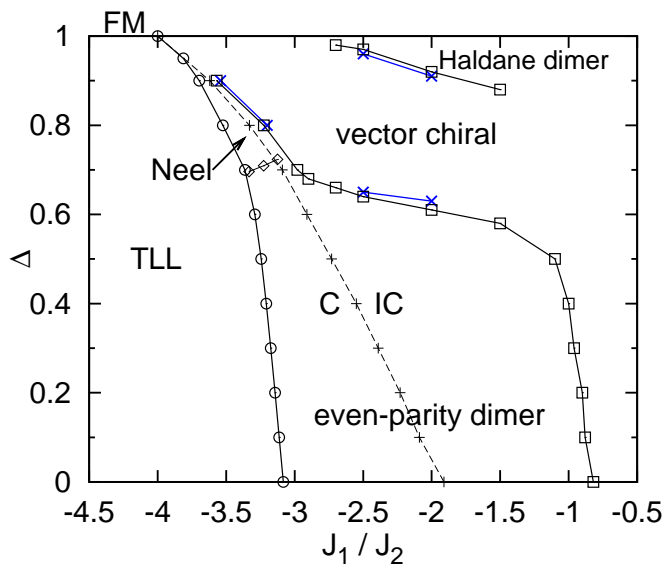}%{phase_jd.eps}
\end{center}
\caption{ (Color online) The ground-state phase diagram of the model
\eqref{eq:H} with $J_1<0$ and $J_2>0$.  The vector chiral phase, which
has a non-vanishing vector chirality \eqref{eq:chirality}, extends
between the two boundaries with the ``$\square$'' symbols.
These boundaries are determined as in Figs.~\ref{fig:order_jm2}
and \ref{fig:order_z0p8}
(see the vertical solid lines in these figures).  Around the highly
degenerate point $(J_1/J_2, \Delta)=(-4,1)$, these two boundaries
could not be determined accurately, but we expect both of them to
continue to this point.
It was also difficult to draw the boundary around the right top corner
of the phase diagram.
The onsets of (Haldane and even-parity) dimer and N\'eel orders occur
inside the vector chiral phase, as
indicated by the ``$\times$'' symbols (determined as in
Figs.~\ref{fig:Cpm_jm2} and \ref{fig:Cpm_z0p8jm3}).  Thus there are
narrow intermediate phases (between the ``$\square$'' and ``$\times$''
symbols) where two kinds of orders coexist.  The phase boundaries
among the TLL, even-parity dimer, and N\'eel phases are determined in
a previous work.\cite{FSF10}
On the right of the ``$\bigcirc$'' symbols, the
even-parity dimer and N\'eel phases alternately appear when
approaching the point $(J_1/J_2, \Delta)=(-4,1)$:
the first transition occurs at $\Delta\approx 0.7$
(``$\diamond$'' symbols) and the second at
$\Delta\approx 0.93$ (not shown).\cite{FSF10}
On the line where $\Delta=1$ and $J_1/J_2<-4$,
the ground state is ferromagnetic (FM).
The ``+'' symbols inside the dimer and N\'eel phases indicate the
Lifshitz line, on which the short-range spin correlation changes its
character from incommensurate (IC) to commensurate (C); see
Fig.~\ref{fig:pitch}.  }
\label{fig:phase}
\end{figure}
%============================

%--------------------------------------
% Short introduction of the paper

In this paper, we study the ground-state properties of the
spin-$\frac12$ frustrated ferromagnetic XXZ chain \eqref{eq:H} with
$J_1<0$ and $J_2>0$, by means of the infinite time evolving block
decimation algorithm (iTEBD)\cite{Vidal07} and effective field
theories based on the bosonization methods.  Previous works on the case
with easy-plane anisotropy $0\le \Delta \le1$ have discussed the
competition among the vector chiral phase with gapless excitations and
the dimer and N\'eel phases with gapped excitations.
\cite{Tonegawa90,Somma01,Chubukov91,Nersesyan98,FSSO08,Sirker10,Itoi01,FSO10,FSF10,SFOF11}
The main goal of this paper is to present a conclusive phase diagram
of the model \eqref{eq:H}, which is shown in Fig.~\ref{fig:phase},
through detailed analyses that extends our previous
works.\cite{FSO10,FSF10,SFOF11} Firstly, we uncover the nature of the
nonmagnetic phase around the SU(2)-symmetric case $\Delta=1$, which
has long been controversial.  We show that this phase has a dimer
order associated with an emergent spin-$1$ degree of freedom on every
other bond.  We term this new phase the Haldane dimer phase.
Secondly, we show the existence of narrow gapped phases that
intervene between the gapless chiral phase and any one of gapped dimer
and N\'eel phases.
As weak inter-chain couplings are turned on,
while the gapless chiral phase evolves into a spiral magnetic order, 
the Haldane dimer
phase can be stabilized by a coupling with phonons due to the
spin-Peierls mechanism.  Our phase diagram may thus provide a useful
starting point for understanding the competing phases in
quasi-1D cuprates.

%--------------------------------------
% Previous results

Let us briefly review previous results on the model \eqref{eq:H} and
summarize our new findings.
While we are mainly concerned with the case of $J_1<0$ and $J_2>0$
in this paper, for comparison,
we also review established results on the case of antiferromagnetic
$J_1,J_2>0$ alongside.

%--------------------------------------
% Classical spin case

In the classical limit $S\to\infty$, the ground state phase diagram of
Eq.~\eqref{eq:H} does not depend on $\Delta$ in the range $0\le
\Delta\le 1$.
The ground state has ferromagnetic order for $J_1/J_2<-4$ and
antiferromagnetic (N\'eel) order for $J_1/J_2>4$.
For $0<|J_1|/J_2<4$, the ground state is in a spiral
magnetic ordered phase, in which the spins rotate by an
incommensurate pitch angle $Q=\pm \arccos (-J_1/4J_2)$ along
the spin chain.  Except for the isotropic case $\Delta=1$, the spiral
plane is fixed in the $xy$ plane, and the vector chirality
\begin{equation} 
 \kappa^z_{\ell,\ell+1} := 
\langle (\bm{S}_\ell \times \bm{S}_{\ell+1})^z \rangle
\label{eq:chirality}
\end{equation}
has a non-vanishing uniform value $\kappa^z_{\ell,\ell+1} = \pm \sin Q$
independent of $\ell$.
Here $\langle\cdots\rangle$ stands for average in the
ground state (with long-range order, if any).

%--------------------------------------
% Quantum spin-1/2 case

In the ground state of the quantum spin-$\frac12$ model, a long-range
magnetic order with broken U(1) spin rotational symmetry is
generally prohibited, unless the uniform magnetic susceptibility is
divergent as in the case of ferromagnetism.\cite{Momoi96} However, a
long-range order (LRO) of the vector chirality
$\kappa^z_{\ell,\ell+1}$ that breaks only the $\mathbb{Z}_2$ parity
symmetry can survive quantum fluctuations in the case of $\Delta\ne
1$.  Using the bosonization theory for $|J_1|/J_2\ll 1$ and $0\le
\Delta <1$, Nersesyan {\it et al.}\cite{Nersesyan98} predicted the
appearance of the vector chiral phase with gapless excitations (as
reviewed in Sec.~\ref{sec:bos_gapless_chiral}).  This gapless chiral
phase shows the spatially uniform vector chirality
$\kappa_{\ell,\ell+1}^z\ne 0$ and power-law decaying (incommensurate)
spiral spin correlations; this phase may therefore be viewed as a
quantum counterpart of the classical spiral phase.  The gapless chiral
phase competes with other quantum phases, in particular, valence bond
solids driven by quantum fluctuations.  In fact, for antiferromagnetic
$J_1>0$, a dimerized phase, in which the singlet state
$(\ket{\ua\da}-\ket{\da\ua})/\sqrt{2}$ (written in the $\{S_\ell^z\}$ basis) is formed on dimerized bonds, 
appears in a large part of the
classical spiral regime $0<J_1/J_2<4$,\cite{Majumdar69, Haldane82,
Nomura94, White96} and the gapless chiral phase appears
only in a small
region in the space spanned by $J_1/J_2$ and
$\Delta$.\cite{Hikihara01}

%--------------------------------------
% Previous works on Ferromagnetic J1<0 case

The phase diagram for the case of ferromagnetic $J_1<0$ and easy-plane
anisotropy $0\le \Delta \le 1$ is presented in Fig.~\ref{fig:phase}.
Early works\cite{Tonegawa90, Somma01} mainly discussed
the transition from the Tomonaga-Luttinger liquid (TLL) phase
to a dimer phase with an even-parity unit\cite{Chubukov91}
$\ket{\uparrow\downarrow}+\ket{\downarrow\uparrow}$ appearing for
$0<\Delta\lesssim 0.7$.  Our recent works\cite{FSO10,FSF10} have
uncovered a rich phase structure in an extended parameter 
space of $J_1/J_2$ and $\Delta$.
In Ref.~\onlinecite{FSO10}, it was shown that the gapless
chiral phase appears in a wide region, and survives up to the close
vicinity of the isotropic case $\Delta=1$ for $-4<J_1/J_2\lesssim
-2.5$ (we also refer to Refs.~\onlinecite{FSSO08,Sirker10} for related
earlier works).  This remarkable stability of the gapless chiral phase
for $J_1<0$ indicates that the sign of $J_1$ plays a crucial role in
stabilizing the vector chirality and the associated ferroelectric
polarization in multiferroic
cuprates.\cite{Masuda05,Park07,Enderle05,Naito07,Yasui11} In
Ref.~\onlinecite{FSF10}, the instability of the TLL phase toward
gapped phases was analyzed using the effective sine-Gordon theory
combined with numerical diagonalization.  It was found that the
even-parity dimer phase\cite{Comment_dimer} discussed in
Refs.~\onlinecite{Tonegawa90,Somma01,Chubukov91} and a N\'eel ordered
phase appear alternately as $\Delta$ is increased on the
right side of the TLL phase in the phase diagram.

%--------------------------------------
% Our work: SU(2) case

An important result of this paper is concerned with 
the nature of the nonmagnetic
phase for $-4<J_1/J_2<0$ around the SU(2)-symmetric case $\Delta=1$.
Previous field-theoretical analyses\cite{Nersesyan98,Itoi01} have
suggested that a dimer phase with a very small
energy gap should appear in this region
(as reviewed in Sec.~\ref{subsec:isotropic_fieldtheory}).
However, neither a dimer order nor an energy gap has been detected in
previous numerical studies.  Using the iTEBD, which 
allow us to treat infinite-size systems directly,
we present the first numerical evidence of a
finite dimer order parameter
\begin{equation}
D_{\ell,\ell+1,\ell+2}= \langle \Svec_\ell \cdot \Svec_{\ell+1} \rangle 
- \langle \Svec_{\ell+1} \cdot \Svec_{\ell+2} \rangle. %\ne 0
\label{eq:dimer}
\end{equation}
Remarkably, this dimer order is associated with ferromagnetic correlations 
$\langle \Svec_\ell \cdot \Svec_{\ell+1} \rangle >0$ of alternating strengths, 
in contrast to antiferromagnetic correlations in singlet dimers for $J_1>0$. 
In this case, it is natural to interpret that effective spin-$1$ degrees of
freedom emerge on the bonds with stronger ferromagnetic
correlation, forming a valence bond solid state\cite{Affleck87}
as in the Haldane chain.\cite{Haldane83} 
We thus call this new phase the Haldane dimer phase. 

%--------------------------------------
% Our work: anisotropic case

We also present detailed analyses of the anisotropic case $\Delta\ne
1$, extending our previous works.\cite{FSO10,FSF10} In particular, we
analyze the transition from the gapless chiral phase to each of the
Haldane dimer, even-parity dimer, and N\'eel phases, and identify
narrow intermediate gapped phases where two kinds of orders coexist
(the regions between ``$\square$'' and ``$\times$'' symbols
in Fig.~\ref{fig:phase}).  Furthermore, we describe how the properties of
various phases can be captured in the language of the Abelian
bosonization\cite{Gogolin98,Giamarchi04} for $|J_1|/J_2 \ll 1$, as
summarized in Table \ref{table:bos}.

%--------------------------------------
% Organization of the paper

The rest of the paper is organized as follows.  In
Sec.~\ref{sec:phases}, we present the numerical results on the order
parameters and half-chain entanglement entropy, which provide the most
basic information for identifying symmetry-broken phases.  In
Sec.~\ref{sec:dimer}, we discuss in detail the dimer phases in the
SU(2)-symmetric case $\Delta=1$ from both
field-theoretical~\cite{Affleck88_review,CFT96,Gogolin98,Giamarchi04}
and numerical analyses.  In Sec.~\ref{sec:easy-plane}, we discuss the
case with the easy-plane anisotropy $0\le\Delta<1$.  In particular, we
review the effective field theory for the gapless chiral
phase,\cite{Nersesyan98} and, following
Ref.~\onlinecite{Lecheminant01}, discuss its instability towards
gapped chiral phases.  The ranges of the gapped chiral phases are then
determined numerically by analyzing the spin correlations.  In
Sec.~\ref{sec:easy-axis}, we briefly describe how the quantum phases
in the easy-axis case\cite{Tonegawa90,Igarashi89_FM,Igarashi89_AFM}
can be understood in the Abelian bosonization framework.  In
Sec.~\ref{sec:conclusions}, we conclude the paper and discuss
implications of our results for quasi-1D cuprates.

%%%%%%%%%%%%%%%%%%%%%%%%%%%%%%%%%%%%%%%%%%%%%%%%%
\section{Numerical analysis of order parameters} \label{sec:phases}
%%%%%%%%%%%%%%%%%%%%%%%%%%%%%%%%%%%%%%%%%%%%%%%%%

%--------------------------------------
% Section introduction

In this section, we present numerical results on several order
parameters and half-chain entanglement entropy calculated by iTEBD.
The vector chiral order parameter and the entanglement entropy are
used to determine the boundaries of the region where
the long-range vector chiral order exists
(the ``$\square$'' symbols in Fig.~\ref{fig:phase}).
The numerical results in this section also suggest the existence of
the narrow intermediate phases (between ``$\square$''
and ``$\times$'' symbols) in which the vector chiral
order coexists with the dimer or N\'eel order.  The precise ranges of
these intermediate phases, however, will be determined in
Sec.~\ref{subsec:gapped_chiral}.

%--------------------------------------
% Method

Before presenting the numerical results, let us briefly
note characteristic features of our numerical method;
for more detailed account of the method, see Supplementary Material of
Ref.~\onlinecite{FSO10}.  The iTEBD algorithm\cite{Vidal07} we
employed is based on the periodic matrix product representation of
many-body wave functions of an infinite system.
It can directly address physical quantities in the thermodynamic limit,
and is free from finite-size or boundary effects.
The (variational) wave function is optimized
to minimize the energy. 
The precision of the algorithm is
controlled by the Schmidt rank $\chi$, which gives the linear
dimension of the matrices.  We exploited the conservation of the total
magnetization $\sum_\ell S^z_\ell=0$ to achieve higher efficiency and
precision of the calculations.  When this algorithm is used in ordered
phases, a variational state finally converges to a symmetry-broken
state with an associated finite order parameter (if it is allowed by
the periodicity of the matrix product state).\cite{note:iTEBD}
In our implementation,
we used a period-4 structure for the variational matrix product state.
In this setting, the vector chiral, dimer, and N\'eel order parameters
analyzed in this section can all be calculated through local
quantities.  In order to allow a finite vector chiral order parameter,
the initial state must contain complex elements as a ``seed'' for the
symmetry breaking.\cite{Okunishi08}

\newcommand{\vN}{\mathrm{vN}}
%============================
\begin{figure}
\begin{center}
 \includegraphics[width=0.48\textwidth]{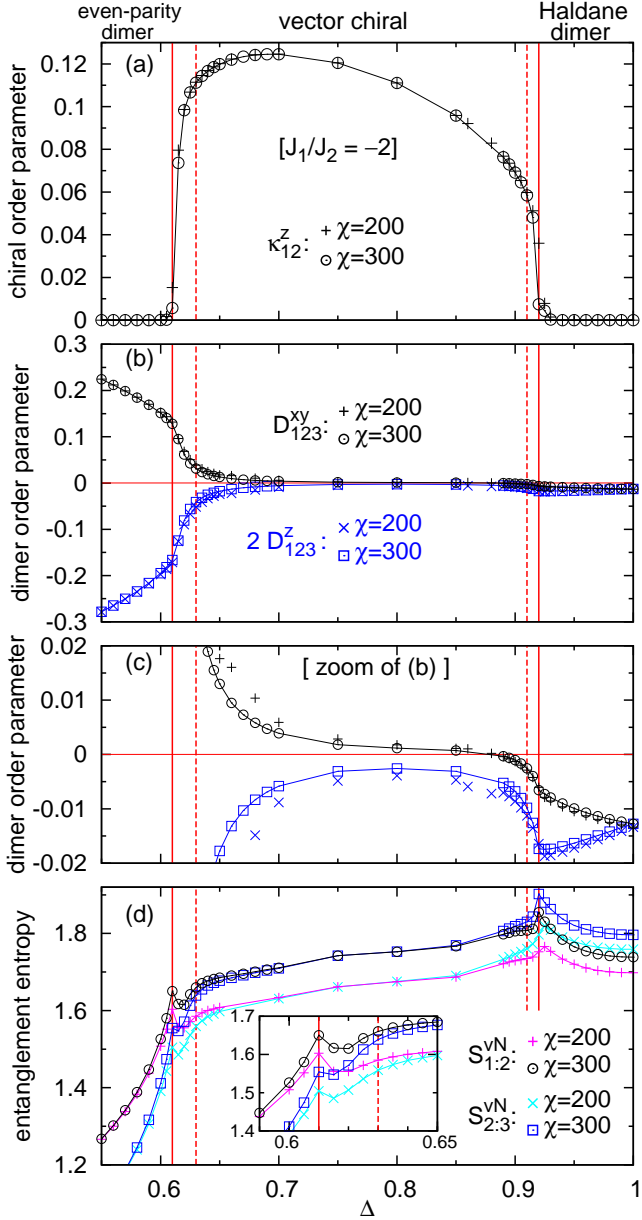}\\%{order_jm2.eps}
\end{center}
\caption{(Color online) (a) Chiral and (b,c) dimer order parameters
and (d) half-chain entanglement entropy as functions of $\Delta$ for
fixed $J_1/J_2=-2$.  These are calculated by the iTEBD with Schmidt
ranks $\chi=200$ and $300$.  Panel (c) is a zoom of panel (b).  In
panel (d), $S^\vN_{1:2}$ and $S^\vN_{2:3}$ are defined for the
bipartitions of the system at the bonds $(1,2)$ and $(2,3)$,
respectively.  Solid vertical lines indicate the boundaries of the
vector chiral phase, and are determined from the onsets of the vector
chiral order parameter in panel (a) or more accurately from the peaks
in the entanglement entropy in panel (d).  Broken vertical lines
indicate the transition points on which dimer orders set in.  These
points are difficult to locate within the analysis of dimer order
parameters in panels (b,c), and are instead determined by the analysis
of spin correlations in Fig.~\ref{fig:Cpm_jm2}.  Narrow intermediate
phases exist between solid and broken vertical lines, where the vector
chiral and dimer orders coexist.  In the intermediate phase in
$0.61\lesssim \Delta\lesssim 0.63$, a dip in the entanglement entropy
is seen in panel (d), as zoomed in the inset.  }
\label{fig:order_jm2}
\end{figure}
%============================

%============================
\begin{figure}
\begin{center}
 \includegraphics[width=0.48\textwidth]{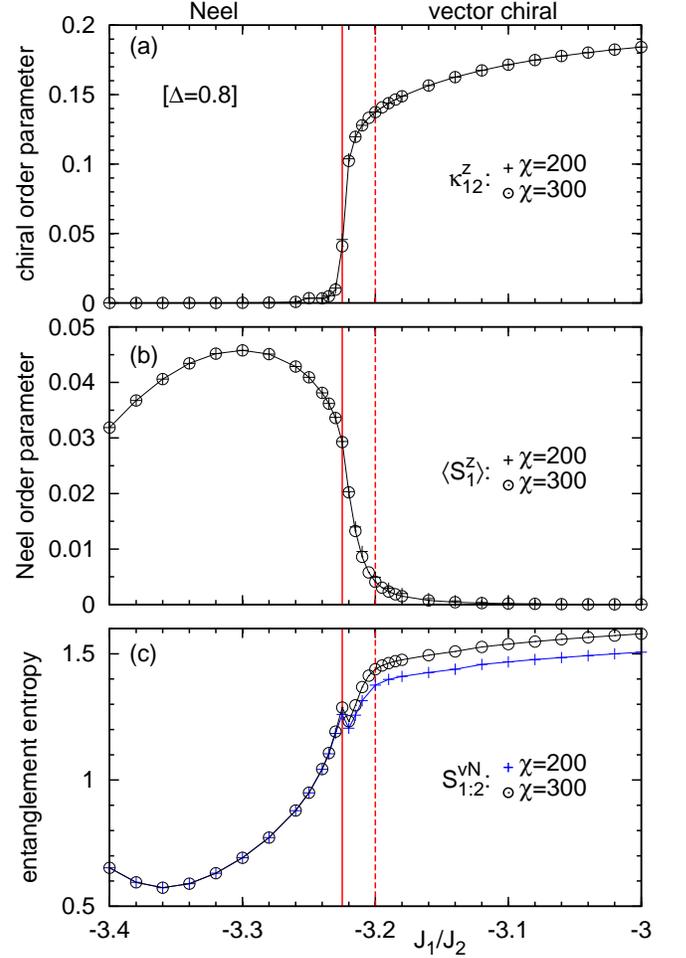}\\%{order_z0p8.eps}
\end{center}
\vspace*{-6mm}
\caption{(Color online) (a) Chiral and (b) N\'eel order parameters and
(c) half-chain entanglement entropy as a function of $J_1/J_2$ for
fixed $\Delta=0.8$, calculated by the iTEBD.  The solid and broken
vertical lines indicate the onsets of the vector chiral and N\'eel
orders, respectively.  The former is determined by the peak position
in the entanglement entropy in panel (c), while the latter is
determined in Fig.~\ref{fig:Cpm_z0p8jm3}.  In the narrow intermediate
phase in $-3.225\lesssim J_1/J_2\lesssim -3.200$, the vector chiral
and N\'eel orders coexist.  }
\label{fig:order_z0p8}
\end{figure}
%============================

%************************************************
\subsection{Vector chiral order}
%************************************************

Figures~\ref{fig:order_jm2} and \ref{fig:order_z0p8} present our
numerical results along the vertical line $J_1/J_2=-2$ and the
horizontal line $\Delta=0.8$, respectively, in the phase diagram (Fig.~\ref{fig:phase}).
Let us first look at the
vector chiral order parameter $\kappa_{12}^z=\langle
(\Svec_1\times\Svec_2)^z \rangle$ displayed in
Figs.~\ref{fig:order_jm2}(a) and \ref{fig:order_z0p8}(a).  This order
parameter is always found to be spatially uniform along the spin chain
in the present model, so we have fixed the site labels.  By observing
the rapid increase of $\kappa_{12}^z$, we find the onset of the vector
chiral phase.  It is natural to think that this rapid increase comes
from the Ising nature of the transition with exponent $\beta=1/8$ for
the spontaneous order parameter, as previously demonstrated in the XY
case $\Delta=0$.\cite{Okunishi08} To determine the transition points
more precisely, however, we use the half-chain entanglement entropy
explained next.

%--------------------------------------
% Half-chain entanglement entropy

The half-chain von Neumann (vN) entanglement entropy is
defined as\cite{Vidal07}
\begin{equation}
 S^\vN = - \sum_{\alpha=1}^\chi \lambda_\alpha^2 \ln \lambda_\alpha^2,
\end{equation}
where $\{\lambda_\alpha\}$ is a set of Schmidt coefficients 
associated with the decomposition of the infinite system into the 
left and right halves and $\chi$ is the Schmidt rank.
As the system approaches a critical point characterized by 
a conformal field theory with a central charge $c$, 
this quantity is known to diverge as\cite{Vidal03,Calabrese04}
\begin{equation}
 S^\vN =\frac{c}{6} \ln \xi + s_1,
\end{equation}
where $\xi$ is the correlation length and $s_1$ is a non-universal
constant.  In an iTEBD calculation with a finite Schmidt rank $\chi$,
the divergence of $S^\vN$ at the critical point is replaced by the
increasing function of $\chi$,\cite{Pollmann09}
\begin{equation} \label{eq:ent_chi}
 S^\vN =\frac{1}{\sqrt{12/c}+1} \ln \chi + s_1', 
\end{equation}
where $s_1'$ is another non-universal constant.  The calculated
entanglement entropy is shown in Figs.~\ref{fig:order_jm2}(d) and
\ref{fig:order_z0p8}(c).  In Fig.~\ref{fig:order_jm2}(d), we plot two
entropies $S^\vN_{1:2}$ and $S^\vN_{2:3}$ associated with the
bipartitions of the system at the bonds $(1,2)$ and $(2,3)$, since
these bonds are inequivalent in the neighboring dimer phases.  By
finding peaks of $S^\vN$, we can determine the boundaries of the
vector chiral phase, more accurately than by using $\kappa_{12}^z$;
see the solid vertical lines in Figs.~\ref{fig:order_jm2} and
\ref{fig:order_z0p8}.  In this way, we have determined the square
symbols in Fig.~\ref{fig:phase}.  Although we could not extract $c$
from the current data of $S^\vN$ using Eq.~\eqref{eq:ent_chi} (which
is expected to be satisfied for larger $\chi$), it is natural to
expect that these critical points are characterized by the
two-dimensional Ising universality class with $c=1/2$ (we again note
that the critical exponent $\beta=1/8$ for this class was confirmed in
the XY case\cite{Okunishi08}).

In most part of the vector chiral phase, the entanglement entropy
increases as a function of $\chi$, indicating a critical nature.
Indeed, in the effective field theory of Nersesyan {\it et
  al.},\cite{Nersesyan98} the gapless chiral phase has $c=1$, and the
increase of $S$ from the cases of $\chi=200$ to $300$ is roughly
consistent with $\Delta S=0.224 \ln(300/200)= 0.091$ expected from
Eq.~\eqref{eq:ent_chi} for $c=1$.  Near the boundaries (solid vertical
lines), the entanglement entropy shows dips,
whose implications will be discussed later.

%[Memo]
% 1/(sqrt{12/c}+1) ln(300/200)= 0.170*0.405 = 0.0687 for c=1/2
%                                         0.224*0.405 = 0.0908 for c=1

%************************************************
\subsection{Dimer orders}
%************************************************

%--------------------------------------
% Dimer order parameters

Next we look at the $xy$ and $z$ components of dimer order parameters,
\begin{subequations}
\begin{align}
 D_{\ell,\ell+1,\ell+2}^{xy} :=
 & \langle (S_\ell^xS_{\ell+1}^x + S_\ell^yS_{\ell+1}^y)  \nonumber\\
 &- (S_{\ell+1}^xS_{\ell+2}^x + S_{\ell+1}^yS_{\ell+2}^y) \rangle, \\
 D_{\ell,\ell+1,\ell+2}^{z} :=
 & \langle S_\ell^zS_{\ell+1}^z - S_{\ell+1}^zS_{\ell+2}^z \rangle. 
\end{align}
\end{subequations}
The alternation of the sign of $D_{\ell,\ell+1,\ell+2}^{xy}$ or
$D_{\ell,\ell+1,\ell+2}^z$ 
along the spin chain would indicate some sort of dimer ordering. 
We assign the site labels in such a way that $D_{123}^z<0$. 
The two order parameters are plotted in Figs.~\ref{fig:order_jm2}(b,c). 
We find that $D_{123}^{xy}$ and $D_{123}^z$ are both finite and have mutually
opposite signs for $\Delta\lesssim 0.65$. 
By contrast, the two order parameters have small finite
values of the same sign for $\Delta\gtrsim 0.9$; 
in spite of the smallness, they are rather stable when the Schmidt rank
$\chi$ is increased as seen in the zoomed plot in Fig.~\ref{fig:order_jm2}(c). 
These results indicate that the dimer phases in the
two regions are of distinct types.

%--------------------------------------
% Even-parity dimer phase

The nature of the dimer phase for $\Delta \lesssim 0.6$ 
can be easily understood as follows.\cite{Chubukov91,FSF10} 
In the XY limit $\Delta=0$, the sign of $J_1$ in Eq.~\eqref{eq:H} 
can be reversed by performing the $\pi$ rotations of spins 
around the $z$ axis on every second sites. 
From the fact that the doubly degenerate ground states
at $(J_1/J_2, \Delta)=(2,0)$ 
are given by the products of singlet dimers, 
one finds, through the above $\pi$-rotation transformation, 
that the exact ground states at $(J_1/J_2, \Delta)=(-2,0)$ are given 
by the dimer states 
whose unit is now replaced by $(\ket{\ua\da}+\ket{\da\ua})/\sqrt{2}$ 
(written in the $\{ S^z_\ell \}$ basis). 
We note that this unit has the even parity with respect to 
the inversion about a bond center, 
in contrast to the odd parity of the singlet dimer at $J_1>0$. 
The direct product states of even-parity dimers show 
$D_{123}^{xy}= -2 D_{123}^z=\pm 1/2$. 
The mutually opposite signs of $D_{123}^{xy}$ and $D_{123}^z$ 
and the approximate relation $D_{123}^{xy}\approx -2 D_{123}^z$
found for $\Delta\lesssim 0.6$ in Fig.~\ref{fig:order_jm2}(b) 
indicate that the even-parity nature of the dimer unit persists
in this region. 
We thus call this phase the even-parity dimer phase.\cite{Comment_dimer} 
It is distinct from the singlet dimer phase appearing for $J_1>0$, 
in which $D_{123}^{xy}$ and $D_{123}^z$ show the same sign. 

%--------------------------------------
% Haldane dimer phase

In the region $\Delta\gtrsim 0.9$ in Fig.~\ref{fig:order_jm2},
$D_{123}^{xy}$ and $D_{123}^z$ are both negative as in the singlet
dimer phase.  However, forming nearest-neighbor singlet dimers is
unlikely for ferromagnetic $J_1<0$.  In Sec.~\ref{sec:dimer}, we point
out that the dimer order in this region is associated with
ferromagnetic nearest-neighbor correlations $\langle
\Svec_\ell\cdot\Svec_{\ell+1} \rangle>0$ of alternating strengths
along the chain, in marked contrast to an antiferromagnetic
correlation in a singlet dimer.  A more detailed comparison of the
dimer phases for $J_1<0$ and $J_1>0$ in the isotropic case
($\Delta=1$) will be presented in Sec.~\ref{sec:dimer}.

%--------------------------------------
% Chiral dimer phase

In the region of a finite vector chiral order
($0.61\lesssim\Delta\lesssim0.92$) in Fig.~\ref{fig:order_jm2}, 
we find that the two dimer order parameters remain finite
in the narrow regions
between the solid and broken vertical lines. 
This indicates the existence of the chiral dimer phases 
(originally predicted in Ref.~\onlinecite{Lecheminant01}), 
in which the vector chiral and dimer orders coexist 
and there are four-fold degenerate ground states below an excitation gap. 
In the entanglement entropy, a dip is seen
in the interval $0.61\lesssim \Delta\lesssim 0.63$, 
which also supports the existence of an intermediate gapped phase.
The peaks in the entanglement entropy indicated by the solid lines
in Fig.~\ref{fig:order_jm2}(d) 
correspond to the Ising critical point between two gapped phases. 
Between the two broken lines in Fig.~\ref{fig:order_jm2}, the dimer order parameters
diminish and the entanglement entropy increases 
as we increase the Schmidt rank $\chi$;
these features are consistent with
the gapless chiral phase. 
The precise determination of the phase boundaries 
between gapped and gapless chiral phases 
is difficult within the analysis of the order parameters and entanglement
entropy in Fig.~\ref{fig:order_jm2}; 
it will be done instead by analyzing spin correlation functions
in Fig.~\ref{fig:Cpm_jm2} in Sec.~\ref{subsec:gapped_chiral}. 

%************************************************
\subsection{N\'eel order}
%************************************************

The appearance of a N\'eel phase with spontaneous staggered
magnetizations $\langle S_\ell^z \rangle \propto (-1)^\ell$ is
discussed in detail in Ref.~\onlinecite{FSF10}.
In Fig.~\ref{fig:order_z0p8}(b), this N\'eel order is detected in the
region $J_1/J_2 \lesssim -3.2$ by measuring $\langle S_1^z \rangle$.
As in the case of the dimer phases, even in the region
where the vector chiral order is finite ($J_1/J_2\gtrsim-3.225$),
the N\'eel order parameter remains finite.
This indicates the existence of a narrow chiral
N\'eel phase, in which the vector chiral and N\'eel orders coexist.
The ground states in this phase should be four-fold degenerate
with a finite excitation gap.
In Fig.~\ref{fig:order_z0p8}(c), a dip in the entanglement entropy
can be found in this region,
consistent with the expected gapped excitation spectrum. 
The precise determination of the transition
point will be done in Fig.~\ref{fig:Cpm_z0p8jm3} in
Sec.~\ref{subsec:gapped_chiral}.

%%%%%%%%%%%%%%%%%%%%%%%%%%%%%%%%%%%%%%%%%%%%%%%%%
\section{Isotropic case $\Delta=1$} \label{sec:dimer}
%%%%%%%%%%%%%%%%%%%%%%%%%%%%%%%%%%%%%%%%%%%%%%%%%

%--------------------------------------
%- Subsection introduction

In this section, we present detailed analyses of the model
\eqref{eq:H} in the isotropic case $\Delta=1$.
While it is known that the singlet dimer phase appears for
$0<J_1/J_2\lesssim 4.15$,\cite{Majumdar69,Haldane82,Nomura94,White96,Eggert96}
the nature of the nonmagnetic ground state in $-4<J_1/J_2<0$ has not
been well understood.
In Sec.~\ref{subsec:isotropic_fieldtheory}, we summarize
previous field-theoretical analyses\cite{Nersesyan98, Itoi01} for
the weak-coupling limit
$|J_1|\ll J_2$, which predicted the appearance of dimer
orders for both signs of $J_1$.
At first glance, this result may seem bizarre since the
singlet dimerization on the $J_1$ bonds, as formed 
in the case of antiferromagnetic $J_1>0$,
is unlikely to occur in the case of
ferromagnetic $J_1<0$.  In Sec.~\ref{subsec:isotropic_numerical}, we
present our numerical results and point out a remarkable difference
between the $J_1>0$ and $J_1<0$ cases in the way how the system hosts
the dimer order.  This leads us to propose the picture
of the ``Haldane dimer phase''
for the dimer phase with $J_1<0$.
Although the ground-state wave functions are largely different
between the Haldane and singlet dimer phases,
we argue that the two phases in fact share a common hidden order.

%************************************************
\subsection{Field-theoretical analyses} \label{subsec:isotropic_fieldtheory}
%************************************************

%--------------------------------------
%- Subsection introduction
Here we summarize previous field-theoretical
analyses\cite{Nersesyan98, Itoi01,Allen97,Cabra00,Kim08} for $|J_1|\ll J_2$.
In this regime, the model \eqref{eq:H} can be viewed as two
antiferromagnetic Heisenberg spin chains which are weakly coupled by
the zigzag interchain coupling $J_1$ as in Fig.~\ref{fig:zigzag}.  We
apply the Abelian and non-Abelian bosonization techniques to describe
the two chains separately, and then treat the interchain coupling
$J_1$ as a weak perturbation.

%------------------------------------------------
\subsubsection{Non-Abelian bosonization} \label{subsec:isotropic_nonabel_bos}
%------------------------------------------------

We start from the non-Abelian 
bosonization\cite{Affleck88_review,CFT96,Gogolin98} 
description of the isotropic model \eqref{eq:H} with $\Delta=1$, 
and present the renormalization group (RG) analysis to identify (marginally) 
relevant perturbations. 

\begin{figure}
\begin{center}
\includegraphics[width=0.5\textwidth]{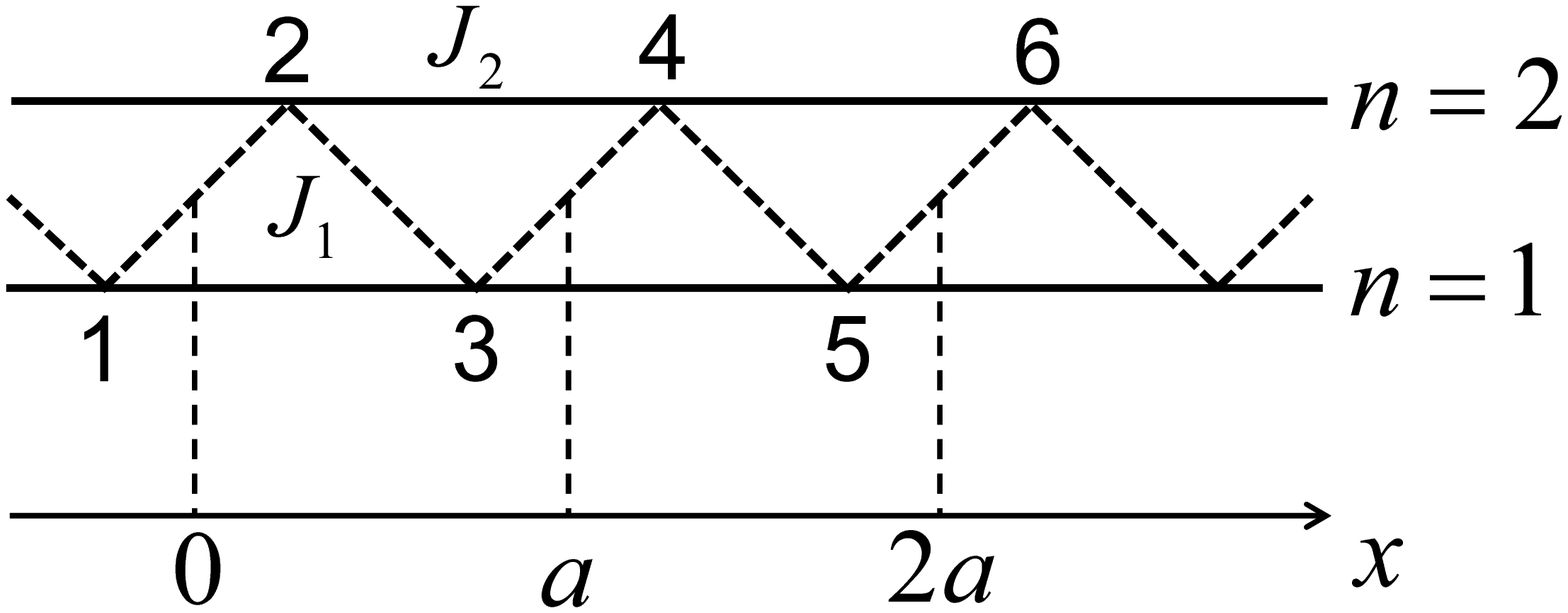}%{zigzag.eps}
\end{center}
\caption{
Zigzag chain picture for the $J_1$-$J_2$ chain model \eqref{eq:H}. 
The $x$ axis indicates the coordinate for the continuum description. 
}
\label{fig:zigzag}
\end{figure}

%--------------------------------------
%- Isolated chain
In the limit $J_1/J_2\to 0$, each isolated antiferromagnetic
Heisenberg chain is described by the SU(2)$_1$ Wess-Zumino-Witten
(WZW) theory, with the spin velocity $v=(\pi/2)J_2a$,
perturbed by a marginally irrelevant backscattering
term.\cite{Affleck88_review,Gogolin98,Eggert96}
The spin operators in the $n$-th chain ($n=1,2$)
can be decomposed as
\begin{equation}
 \Svec_{2j+n} \to a [\Mvec_n (x_n) + (-1)^j \Nvec_n(x_n)]
\end{equation}
with $x_1(j)=(j-\frac14)a$ and $x_2(j)=(j+\frac14)a$,
where $a$ is the lattice spacing of each chain; see Fig.~\ref{fig:zigzag}.
The uniform and staggered components, $\Mvec_n$ and $\Nvec_n$,
have the scaling dimensions $1$ and $1/2$, respectively.
The former can be decomposed
into chiral (right and left) components:
$\Mvec_n=\Mvec_{nR}+\Mvec_{nL}$.  Another important operator is the
(in-chain) staggered dimerization operator $\epsilon_n$ define by
\begin{equation}
   (-1)^j \Svec_{2j+n}\cdot\Svec_{2j+n+2} \to a \epsilon_n (x_n),  
\end{equation}
which has the scaling dimension $1/2$. 

%============================
%============================

%============================
\begin{figure}
\begin{center}
\includegraphics[width=0.48\textwidth]{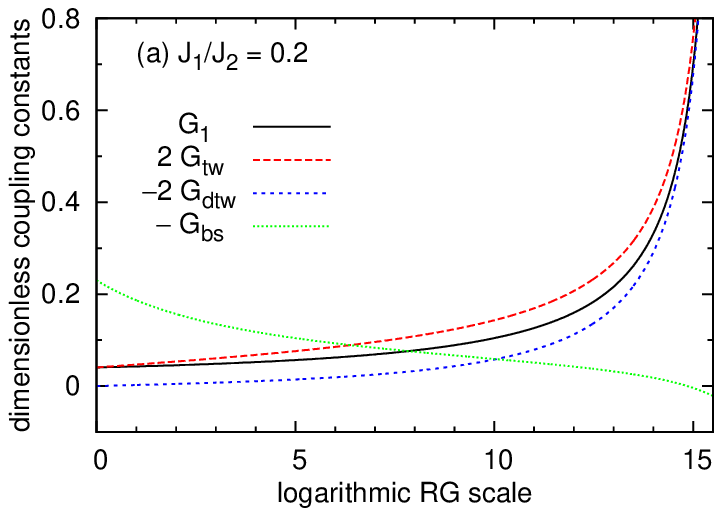}\\%{RG_jp0_2.eps}\\
\includegraphics[width=0.48\textwidth]{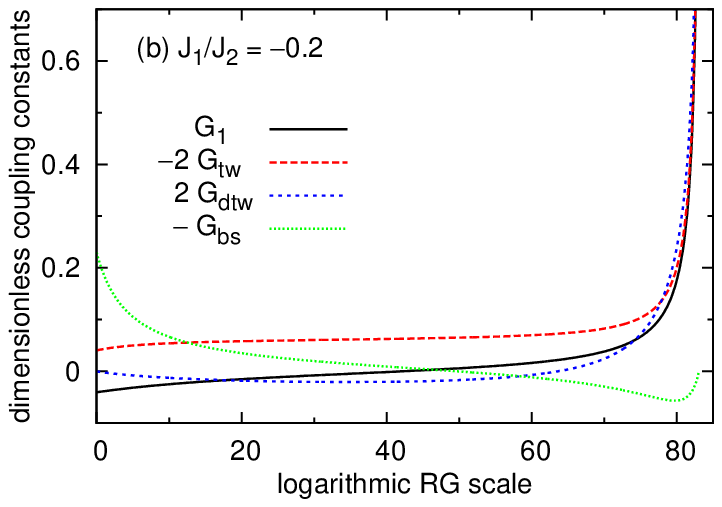}%{RG_jm0_2.eps}
\end{center}
\caption{Numerical solutions to the one-loop RG equations
\eqref{eq:RGeq} for (a) $J_1/J_2=0.2$ and (b) $J_1/J_2=-0.2$.  We set
$\lambda=1$ [see Eq.~\eqref{eq:G_i}].  It is found that the three
coupling constants $G_1$, $G_\tw$, and $G_\dtw$ are
most relevant and grow under the RG, with asymptotically
a simple ratio
$G_1:G_\tw:G_\dtw=2:1:(-1)$ or $2:(-1):1$ for $J_1>0$ and $J_1<0$,
respectively.  In the plots, factors $\pm 2$ are muliplied to $G_\tw$
and $G_\dtw$ so that these ratios can be visually confirmed.  }
\label{fig:RG_flow}
\end{figure}
%============================

%--------------------------------------
%- Coupled chain, RG equations

The inter-chain zigzag coupling $J_1$ produces at most marginal perturbations, 
in the RG sense,  around the WZW fixed point; 
relevant perturbations such as $\Nvec_1\cdot\Nvec_2$ are prohibited by
the symmetry of the zigzag chain model.  
The symmetry-allowed marginal perturbations are summarized as
\begin{equation}
 H' = \int dx \sum_i g_i \Ocal_i,
\end{equation}
where $i$ runs over the following five operators:\cite{Itoi01}
\begin{subequations}\label{eq:O_marginal}
\begin{align}
 &\Ocal_\bs = \Mvec_{1R} \cdot \Mvec_{1L} + \Mvec_{2R} \cdot \Mvec_{2L},\\
 &\Ocal_1 = \Mvec_{1R} \cdot \Mvec_{2L} + \Mvec_{1L} \cdot \Mvec_{2R},\\
 &\Ocal_2 = \Mvec_{1R} \cdot \Mvec_{2R} + \Mvec_{1L} \cdot \Mvec_{2L},\\
 &\Ocal_\tw = \frac{a}2 (\Nvec_1\cdot\partial_x\Nvec_2
                          - \Nvec_2\cdot\partial_x\Nvec_1),\\
 &\Ocal_\dtw = \frac{a}2 (\epsilon_1\partial_x\epsilon_2
                           - \epsilon_2\partial_x\epsilon_1).
\end{align}
\end{subequations}
Here $\Ocal_\bs$ is the backscattering term present in isolated chains.  
The zigzag $J_1$ coupling produces the current-current interactions,
$\Ocal_1$ and $\Ocal_2$, 
and the twist operator $\Ocal_\tw$.  
The dimer twist operator $\Ocal_\dtw$ is generated in the RG process
as we see later. 
The bare coupling constants are given by
\begin{gather}\label{eq:g_bare}
 g_\bs (0)=-0.23(2\pi v), ~~
 g_1 (0)=g_2 (0)=2J_1a,\\
 g_\tw (0)=J_1a,~~g_\dtw (0)=0,
\end{gather}
where $g_\bs(0)$ was estimated in Ref.~\onlinecite{Eggert96}. 
All the operators in Eq.~\eqref{eq:O_marginal} have the scaling
dimensions $2$, and their competition in the RG flow must be
analyzed carefully by deriving the RG equations. 
We define the dimensionless coupling constants
\begin{align}\label{eq:G_i}
 &G_i = \frac{g_i}{2\pi v} ~~(i=\bs,1,2),\\
 &G_i = \frac{g_i}{2\pi v \lambda^2} ~~(i=\tw,\dtw), 
\end{align}
where $\lambda$ is a dimensionless constant
of order unity. 
Using the operator product expansions in the WZW
theory,\cite{CFT96,Shelton96,Starykh04,Starykh05,Hikihara10} 
the one-loop RG equations\cite{Cardy96} are derived
as\cite{Nersesyan98,Itoi01,Cabra00,Kim08}
\begin{subequations}\label{eq:RGeq}
\begin{align}
 &\dot{G}_\bs = G_\bs^2 + G_\tw^2 - G_\dtw^2,\\ 
 &\dot{G}_1 = G_1^2 + G_\tw^2 - G_\tw G_\dtw, \label{eq:RGeq_G1}\\
 &\dot{G}_\tw = -\frac12 G_\bs G_\tw + G_{1} G_\tw -\frac12 G_{1}G_{\dtw},\\
 &\dot{G}_\dtw = \frac32 G_\bs G_\dtw - \frac32 G_{1} G_\tw, 
\end{align}
\end{subequations}
where the dot indicates the derivative ($\dot{G}_i=dG_i/dl$)
with respect to the change of the cutoff: $a\to e^{dl} a$. 
See Appendix \ref{app:RG} for the derivation of Eq.~\eqref{eq:RGeq}. 
We have ignored $G_2$ since it does not affect the flow of
the other coupling constants at the one-loop level.  

%--------------------------------------
%- Numerical solutions to RG equations

Numerical solutions to the RG equations \eqref{eq:RGeq} are presented
in Fig.~\ref{fig:RG_flow}.  For both signs of $J_1$, the three
coupling constants $G_{1}$, $G_\tw$, and $G_\dtw$ finally grow to
large values under the RG;\cite{Nersesyan98,Itoi01} 
they asymptotically have the simple ratio $G_1:G_\tw:G_\dtw=2:1:(-1)$ or
$2:(-1):1$ for $J_1>0$ and $J_1<0$, respectively.  Remarkably, $G_{1}$
finally grows with a {\it positive} sign for both signs of $J_1$.  For
$J_1<0$, in particular, it is initially negative but changes sign
before starting to grow in the RG process.  By contrast, $G_\tw$
retains the same sign as its initial value.
The properties of the fixed points governed by large $G_{1}(>0)$,
$G_\tw$, and $G_\dtw$ are non-trivial.
In fact, while the non-Abelian formalism allows us to
derive the RG equations in a manifestly SU(2)-invariant form, it is
often not very useful for discussing the physical roles
of (marginally) relevant perturbations.
In the next section, we proceed to the
Abelian bosonization analysis to show that the positive development of
$G_1$ induces a gapped state with a finite dimer order parameter
$D_{123}\ne 0$.

%--------------------------------------
%- Large correlation length for J1<0

As seen in Fig.~\ref{fig:RG_flow}(a) and (b), the coupling constants
grow much more slowly for $J_1<0$ than for $J_1>0$.
This implies that for $J_1<0$, the energy gap associated
with the dimer order should be much smaller and
the spin correlation length $\xi$ should be much larger.
In fact, as argued by Itoi and Qin,\cite{Itoi01} the
correlation length becomes of astronomical scale [e.g., $\xi/a\sim
e^{83}\sim 10^{36}$ for the case of Fig.~\ref{fig:RG_flow}(b)].  Such
a tiny gap or a large correlation length is very difficult to detect
by any numerical investigation; the system effectively behaves like a
gapless system even when the system size is macroscopically large.  We
stress, however, that this insight is based on the perturbative RG
analysis for small $J_1/J_2<0$, and it is possible
that the energy gap
grows to an observable magnitude as we increase $|J_1|/J_2$.  Our
numerical result presented in Sec.~\ref{subsec:isotropic_numerical}
indeed identifies a large but detectable correlation lengths around
$J_1/J_2=-2$.

%------------------------------------------------
\subsubsection{Abelian bosonization}\label{subsec:isotropic_abel_bos}
%------------------------------------------------

In this section, we use the Abelian bosonization formalism~\cite{Giamarchi04} 
to discuss the physical roles of the marginally relevant perturbations 
$G_1(>0)$, $G_\tw$, and $G_\dtw$ identified in the non-Abelian analysis. 
Although the Abelian formalism obscures the SU(2) symmetry of the model, 
it has the advantage of simplifying identification of 
various orders with the pattern of locking of
bosonic fields, as illustrated in Table~\ref{table:bos}. 

%--------------------------------------
%- Isolated XXZ chain

Let us start from the two decoupled antiferromagnetic chains in the
limit $J_1/J_2\to 0$.  We summarize the Abelian bosonization
description\cite{Gogolin98,Giamarchi04} of a single XXZ chain 
($0\le \Delta\le 1$), so that the same formulation can be used later in
Sec.~\ref{subsec:bos_easy-plane}.
Each decoupled XXZ chain labeled by
$n=1,2$ is described by a Gaussian Hamiltonian
\begin{equation}\label{eq:H_Gauss_n}
 H_n = \int dx \frac{v}2 \left[ K (\partial_x\theta_n)^2 
+ K^{-1} (\partial_x\phi_n)^2 \right] 
\end{equation}
where the velocity $v$ and the TLL parameter $K$ are given by
\begin{equation}\label{eq;K_v_bos}
 v = \frac{\pi\sqrt{1-\Delta^2}}{2 \arccos \Delta} J_2 a, ~~~
 K = \frac1{1-(1/\pi) \arccos\Delta}. 
\end{equation}
The bosonic fields $\phi_n$ and $\theta_n$ satisfy the commutation relation
\begin{equation}\label{eq:comm_bos}
 [\phi_n (x), \theta_{n'}(x')] = i \delta_{nn'} Y(x-x'),
\end{equation}
where $Y(x-x')$ is the step function  
\begin{equation}
 Y(x-x') = 
 \begin{cases}
 0 & (x<x'), \\
 1/2 & (x=x'), \\
 1 & (x>x').
 \end{cases} 
\end{equation}
The spin and (in-chain) dimer operators are expressed in terms of the
bosonic fields as
\begin{align}
 \label{eq:Sz_bos}
 &S^z_{2j+n}= \frac{a}{\sqtp} \partial_x \phi_n(x_n)
             + (-1)^j A_1 \cos [\sqtp\phi_n(x_n)]+\dots,\\
 \label{eq:Sp_bos}
 &S^+_{2j+n} % \notag\\
 = e^{i\sqtp\theta_n(x_n)}
    \left\{ (-1)^jB_0 \right. \notag\\
 &\left.\qquad\qquad\qquad\qquad\quad
   + B_1 \cos[\sqtp\phi_n(x_n)] +\dots\right\},  \\
 & (-1)^j \Svec_{2j+n}\cdot\Svec_{2j+n+2}  = C \sin (\sqtp\phi_n) + \dots, 
\end{align}
where $A_1$, $B_0$, $B_1$ (Refs.~\onlinecite{Hikihara98,Lukyanov97}), and 
$C$ (Ref.~\onlinecite{Takayoshi10}) are non-universal constants
which depend on $\Delta$.

%--------------------------------------
%- Coupled chains

We now focus on the case $\Delta=1$, at which $K=1$. 
To treat the coupled chains, it is useful to introduce the bosonic fields 
for symmetric $(+)$ and antisymmetric $(-)$ sectors:
\begin{equation}\label{eq:phi_theta_pm}
 \phi_\pm = \frac1{\sqrt{2}} (\phi_1 \pm \phi_2),~~
 \theta_\pm =  \frac1{\sqrt{2}} (\theta_1 \pm \theta_2). 
\end{equation}
The three perturbations found to grow in the non-Abelian analysis 
have the following
expressions:\cite{Nersesyan98,Cabra00,Zarea04,Comment_O1_bos}
\begin{subequations}\label{eq:O_bos}
\begin{align}
\label{eq:O1_bos}
\Ocal_1 =& -\frac{B_1^2}{2a^2} \cos(\sqfp\phi_+) \cos(\sqfp\theta_-) \notag\\
 &+ \frac1{8\pi} \left[ (\partial_x\phi_+)^2 - (\partial_x\theta_+)^2
 -(\partial_x\phi_-)^2 + (\partial_x\theta_-)^2 \right],\\
 \label{eq:Otw_bos}
 \Ocal_\tw =& \frac{\sqp B_0^2}{a} (\partial_x\theta_+) \sin(\sqfp\theta_-)
               \notag\\
 &+ \frac{\sqp A_1^2}{2a} \left[ (\partial_x\phi_+)\sin(\sqfp\phi_-)
  \right. \notag\\
 &\left.\qquad\qquad\quad
 + (\partial_x\phi_-)\sin(\sqfp\phi_+)\right], \\
 \Ocal_\dtw =& \frac{\sqp C^2}{a} \left[ (\partial_x \phi_+)\sin(\sqfp\phi_-)
  \right. \notag\\
 &\left.\qquad\qquad - (\partial_x \phi_-) \sin(\sqfp\phi_+) \right]. 
\end{align}
\end{subequations}
Furthermore, the $\Ocal_2$ term, which is decoupled from the other terms in the RG equation \eqref{eq:RGeq}, has the expression
\begin{equation}\label{eq:O2_bos}
\begin{split}
\Ocal_2 =& -\frac{B_1^2}{2 a^2} \cos(\sqfp\phi_-) \cos(\sqfp\theta_-) \\
&+ \frac{1}{8\pi} \left[ (\partial_x \phi_+)^2 + (\partial_x \theta_+)^2 - (\partial_x \phi_-)^2 - (\partial_x \theta_-)^2 \right] .
\end{split}
\end{equation}
The second lines of Eq.~\eqref{eq:O1_bos} and Eq.~\eqref{eq:O2_bos} can be combined
with the Gaussian Hamiltonians \eqref{eq:H_Gauss_n} of
the decoupled chains, leading to 
\begin{equation}\label{eq:H_Gauss_pm}
 H_0 = \int dx \sum_{\nu=\pm} \frac{v_\nu}2
   \left[ K_\nu (\partial_x\theta_\nu)^2
        + K_\nu^{-1} (\partial_x\phi_\nu)^2 \right]
\end{equation}
with 
\begin{equation}
\begin{split}
 &K_\pm = 1\mp \frac{G_1}2 + O(G_1^2,G_2^2),\\
 &v_\pm = v \left[ 1 \pm \frac{G_2}2 + O(G_1^2,G_2^2) \right].  
\end{split}
\end{equation}
Using the new Gaussian Hamiltonian $H_0$, 
we can calculate the scaling dimension of the operators
in Eq.~\eqref{eq:O_bos}. 
Specifically, the scaling dimension of $e^{i\sqfp\phi_\pm}$
and $e^{i\sqfp\theta_\pm}$ is given
by $K_\pm$ and $K_\pm^{-1}$, respectively. 
In the non-Abelian analysis, we have seen that $G_1$ grows to a positive
value in the RG flow irrespective of the sign of $J_1$. 
Assuming $G_1>0$, we find that the product of the
two cosine operators in the first line of Eq.~\eqref{eq:O1_bos}
(with scaling dimension $2-G_1$) is the most relevant term
among those in Eq.~\eqref{eq:O_bos}. 
This term locks the bosonic fields at 
\begin{equation}\label{eq:lock_dimer}
(\sqfp\phi_+,\sqfp\theta_-)=(0,0) ~~\text{or}~~ (\pi,\pi). 
\end{equation}
These correspond respectively to 
finite positive or negative value of the dimer order parameter 
$D_{123}= \langle \Svec_1 \cdot \Svec_2 \rangle 
- \langle \Svec_2 \cdot \Svec_3 \rangle$, 
since the (inter-chain) dimer operator is expressed as  
\begin{equation}\label{eq:bos_dimer}
\begin{split}
 &\Svec_{2j+1}\cdot\Svec_{2j+2} - \Svec_{2j+2}\cdot\Svec_{2j+3}
 = 2a^2 \Nvec_1\cdot\Nvec_2 +\dots \\
 &\approx 2B_0^2\cos(\sqfp\theta_-) 
   + A_1^2 \left[ \cos(\sqfp\phi_+) + \cos(\sqfp\phi_-) \right] . 
\end{split}
\end{equation}
In the last expression, the first term and the rest come from 
the $xy$ and $z$ components of the spins, respectively. 
For the locking in Eq.~\eqref{eq:lock_dimer}, 
these components acquire both positive or both negative expectation values,  
in agreement with Fig.~\ref{fig:order_jm2}(c) and with 
the SU(2) symmetry of the model. 
It is worth noting that the locking positions of the two degenerate
ground states in Eq.~\eqref{eq:lock_dimer} 
are independent of the sign of $J_1$ in the isotropic case $\Delta=1$. 
The second most relevant terms in Eq.~\eqref{eq:O_bos} are 
$(\partial_x\theta_+) \sin(\sqfp\theta_-)$ and 
$(\partial_x\phi_-)\sin(\sqfp\phi_+)$
with scaling dimension $2-G_1/2$. 
As explained in Sec.~\ref{subsec:bos_easy-plane}, 
the former has the effect of inducing the incommensurability
in spin correlations.\cite{Nersesyan98} 
Since a finite energy gap opens due to $G_1>0$ in the dimer phases, 
the incommensurate spin correlations are expected to remain short-ranged.

%************************************************
\subsection{Numerical results and physical properties of dimer phases} 
\label{subsec:isotropic_numerical}
%************************************************

%--------------------------------------
%- Subsection introduction

In this section, we present numerical results on the model \eqref{eq:H} 
in the isotropic case $\Delta=1$, 
and discuss physical properties of the dimer phases for 
different signs of $J_1$. 
In agreement with the field-theoretical results reviewed 
in the previous section, 
we find that the dimer order parameter $D_{123}$ becomes finite 
for both signs of $J_1$, 
and that there are doubly degenerate ground states with 
positive and negative $D_{123}$. 
While we propose different physical pictures for the dimer orders 
in the $J_1>0$ and $J_1<0$ cases (Sec.~\ref{subsubsec:dimer_corr}), 
we also discuss a hidden order common to the two cases 
(Sec.~\ref{subsubsec:dimer_hidden}). 
In the following, our numerical results (based on iTEBD with $\chi=300$) 
are presented for the ground state with $D_{123}<0$.

%------------------------------------------------
\subsubsection{Local spin correlations} \label{subsubsec:dimer_corr}
%------------------------------------------------

%--------------------------------------
%- Subsection introduction & Antiferromagnetic case

\begin{figure}
\begin{center}
\includegraphics[width=0.48\textwidth]{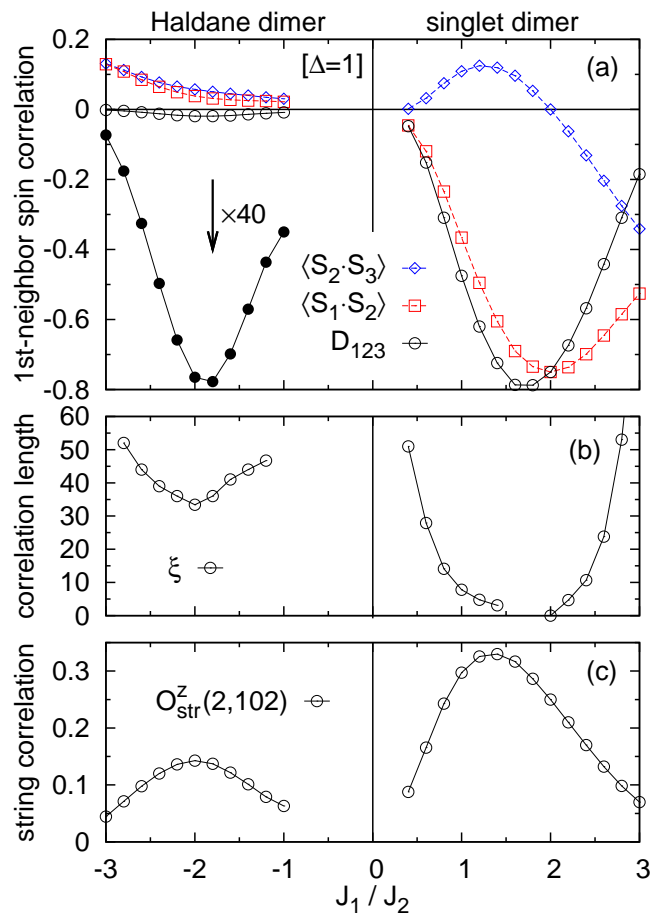}%{dim_z1.eps}
\end{center}
\caption{(Color online)
(a) Nearest-neighbor spin correlations
$\langle \Svec_j \cdot \Svec_{j+1} \rangle$ and
the dimer order parameter $D_{123}$, 
(b) the spin correlation length $\xi$, 
and  
(c) the string correlation \eqref{eq:strg} with $\ell=2$ and $r=50$, 
as a function of $J_1/J_2$ in the isotropic case $\Delta=1$. 
In panel (a), $D_{123}$ multiplied by $40$ is also
plotted for $J_1/J_2<0$ (filled circular symbols).
In panel (b), $\xi$ is too small to determine around $J_1/J_2=2$. 
}
\label{fig:dim_z1}
\end{figure}

In Fig.~\ref{fig:dim_z1}(a), we plot nearest-neighbor spin correlations 
$\langle \Svec_\ell \cdot \Svec_{\ell+1} \rangle$ (with $\ell =1,2$) 
and the dimer order parameter 
$D_{123}= \langle \Svec_1 \cdot \Svec_2 \rangle 
- \langle \Svec_2 \cdot \Svec_3 \rangle$
for $-3\le J_1/J_2 \le 3$. 
While $D_{123}\ne 0$ can be confirmed for both $J_1>0$ and $J_1<0$, 
a notable difference between the two cases can be found in the signs 
of local spin correlations. 

%============================
%============================

%============================
\begin{figure}
\begin{center}
\includegraphics[width=0.48\textwidth]{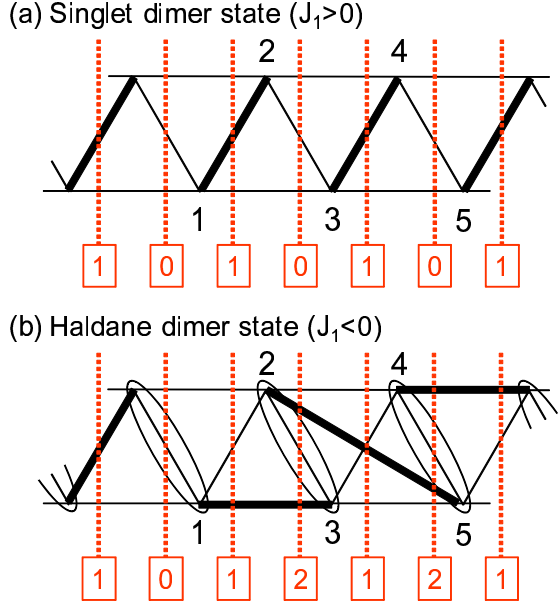}%{dimer.eps}
\end{center}
\caption{(Color online)
Sketches of (a) the singlet dimer state and (b) the Haldane dimer state.  
The thick lines indicate valence bonds. 
In (b),  the encircled bonds indicate emergent spin-$1$'s. 
From each of them, a valence bond emanate to each left and right;   
the wave function is given by a superposition of such valence bond
covering states. 
Vertical cuts (dashed lines) are introduced to probe a hidden order; 
the number of valence bonds crossing with each cut is shown in the square. 
The alternation of odd and even numbers is found in both the states. 
}
\label{fig:dimer}
\end{figure}
%============================

%--------------------------------------
%- Antiferromagnetic case

For $J_1>0$, one of the following inequalities is always satisfied: 
\begin{subequations}\label{eq:corr_sd}
\begin{align}
 &\langle \Svec_1 \cdot \Svec_2 \rangle <
 -\langle \Svec_2 \cdot \Svec_3 \rangle <  0
  \quad (0<J_1/J_2<2), \\
 &\langle \Svec_1 \cdot \Svec_2 \rangle < 
 \langle \Svec_2 \cdot \Svec_3 \rangle \le 0
  \quad (2\le J_1/J_2\lesssim 4.15). 
\end{align}
\end{subequations}
Namely, the system has a strong {\em antiferromagnetic} correlation on
the bond $(1,2)$ and a weaker correlation on $(2,3)$.
In this case, it is natural to assume that singlet dimers
are formed on the bonds $(2j+1,2j+2)~(j\in \mathbb{Z})$,
and are weakly correlated with each other,
as schematically shown in Fig.~\ref{fig:dimer}(a).
Hence we call this phase the singlet dimer phase.
In particular, the ground state is exactly given
by a direct product of singlet dimers at the Majumdar-Ghosh
point\cite{Majumdar69} $J_1/J_2=2$.  In Fig.~\ref{fig:dim_z1}(a) we
find that the weaker correlation $\langle \Svec_2 \cdot \Svec_3
\rangle$ changes the sign at this point.

%--------------------------------------
%- Ferromagnetic case

By contrast, the following inequality is found to be satisfied when $-4<J_1/J_2<0$: 
\begin{equation}\label{eq:corr_hd}
 0< \langle \Svec_1 \cdot \Svec_2 \rangle < 
\langle \Svec_2 \cdot \Svec_3 \rangle. 
\end{equation}
Namely, strong and weak {\em ferromagnetic} nearest-neighbor
correlations alternate along the chain.
This observation led us to propose that there should
be emergent spin-$1$ degrees of freedom on the bonds
$(2j+2,2j+3)~(j\in \mathbb{Z})$ that have stronger ferromagnetic correlation,
as depicted by ellipses in Fig.~\ref{fig:dimer}(b).
Since the total wave function is a spin singlet, such spin-$1$'s
are expected to form a valence bond solid state\cite{Affleck87}
as in the spin-$1$ Haldane chain.\cite{Haldane83}
Namely, from each encircled bond in Fig.~\ref{fig:dimer}(b),
two valence bonds emanate, one to the left and one to the
right; the total wave function is obtained by superposing such valence
bond covering states.
We thus call the dimer phase with $J_1<0$ the
Haldane dimer phase.  The emergence of the Haldane chain physics in
this phase is also supported by the presence of a hidden non-local
order analyzed in Sec.~\ref{subsubsec:dimer_hidden}.

%--------------------------------------
%- Magnitude of the dimer order parameter

In Sec.~\ref{subsec:isotropic_fieldtheory}, it was argued that 
the marginal perturbation $G_1$, which induces the
dimer order, grows very slowly under the RG for $J_1<0$ 
and that the energy gap associated with the dimer order can be
extremely small.\cite{Itoi01} 
The result of Fig.~\ref{fig:dim_z1}(a) indicates that the dimer order
parameter $D_{123}$ grows to a numerically detectable magnitude
for intermediate values of $|J_1|/J_2 \,(\approx 2)$, 
although the obtained values are much much smaller
compared to the $J_1>0$ case (by a factor of around $1/40$). 
The weakness of the effect of $J_1$ in inducing the dimer order and
the associated energy gap for $J_1<0$ 
is also seen in the spin correlation length discussed next.

%------------------------------------------------
\subsubsection{Spin correlation length} \label{subsubsec:dimer_corrlen}
%------------------------------------------------

%============================
\begin{figure}
\begin{center}
\includegraphics[width=0.48\textwidth]{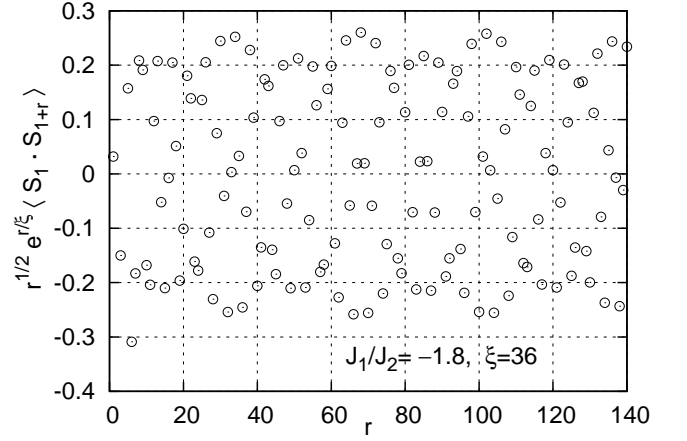}%{Css_z1jm1p8.eps}
\end{center}
\caption{
Determination of the spin correlation length $\xi$, 
illustrated for $J_1/J_2=-1.8$. 
Assuming the asymptotic behavior \eqref{eq:spcorr_dimer}, 
we plot the function $r^{1/2}e^{r/\xi} \langle \Svec_1\cdot\Svec_{1+r} \rangle$, 
and tune $\xi$ such that the oscillation width of this function becomes 
as constant as possible as a function of $r$.\cite{White96} 
While the oscillations arise from the cosine factor in \eqref{eq:spcorr_dimer}, 
it is difficult to extract the pitch angle $Q$ from this figure; 
instead, calculations in Sec.~\ref{subsec:pitch} give $Q/(2\pi) \approx 0.235$. 
}
\label{fig:Css_z1}
\end{figure}
%============================

%--------------------------------------
%- How to determine xi
We determine the spin correlation length $\xi$ in the dimer phases
by using the method of Ref.~\onlinecite{White96}. 
Except at the Lifshitz point $J_1/J_2=2$, 
the spin correlation function is expected to behave 
at long distances as\cite{White96,Nomura05}
\begin{equation}\label{eq:spcorr_dimer}
 \langle \Svec_1\cdot\Svec_{1+r} \rangle \approx
 A \cos (Q r) r^{-\frac12} e^{-r/\xi}. 
\end{equation}
In the incommensurate regions $-4<J_1/J_2<0$ and $0<J_1/J_2<2$, 
the pitch angle $Q$ changes continuously from $0$ to $\pi$,
as will be discussed in Sec.~\ref{subsec:pitch}
(see Fig.~\ref{fig:pitch}). 
For $2<J_1/J_2\lesssim 4.15$, $Q$ is fixed at $Q=\pi$. 
To determine $\xi$, we plot
$r^{1/2}e^{r/\xi} \langle \Svec_1\cdot\Svec_{1+r} \rangle$
as a function of $r$, 
and tune $\xi$ such that the amplitude of oscillations 
becomes as constant as possible, as illustrated in Fig.~\ref{fig:Css_z1}. 
While the coefficient $A$ in Eq.~\eqref{eq:spcorr_dimer} is given by
the oscillation amplitude in Fig.~\ref{fig:Css_z1}, 
it is not simple to determine $Q$ which can fit these very rapid oscillations;
instead 
it will be determined by calculating the spin structure factor
in Fig.~\ref{fig:Spm_z0p8}. 

%--------------------------------------
%- Result of xi

The calculated $\xi$ is plotted in Fig.~\ref{fig:dim_z1}(b). 
The data for $J_1>0$ are broadly in agreement with 
Ref.~\onlinecite{White96}.\cite{Comment_White96} 
We find that the values of $\xi$ are much larger for $J_1<0$ than for $J_1>0$, 
as anticipated from the magnitudes of the dimer order parameter 
in Fig.~\ref{fig:dim_z1}(a). 

%--------------------------------------
%- Spin gap

We use the above numerical data of the spin correlation
length $\xi$ to infer the magnitude of the spin gap $\Delta_s$ for $J_1<0$. 
In general the spin gap $\Delta_s$ should be
inversely proportional to $\xi$, with the proportionality constant
being the spin velocity.
From the data of Ref.~\onlinecite{White96} for $J_1>0$, 
we extract an approximate relation $(\Delta_s/J_2) \xi \approx 2$. 
Applying the same relation to the $J_1<0$ case,  
we estimate the spin gap $\Delta_s$
around $J_1/J_2=-2$ to be roughly equal to $0.06 J_2$.
We note that this should be considered as a crude
order of magnitude estimate.

%------------------------------------------------
\subsubsection{Hidden order}  \label{subsubsec:dimer_hidden}
%------------------------------------------------

%--------------------------------------
%- Hidden order, string correlation function

The singlet and Haldane dimer phases have different
(local) features of short-range correlations as
expressed in Eqs.~\eqref{eq:corr_sd} and \eqref{eq:corr_hd}.
In spite of this local difference, the two phases in fact share a common
non-local order, as we now explain.
Let us count the number of valence bonds crossing
the vertical cuts (dashed lines) depicted in Fig.~\ref{fig:dimer}.
We find that even and odd numbers alternate in the same way in the two
phases, when we take the ground state with $D_{123}<0$.
The existence of such a hidden non-local order can be probed numerically
by calculating the string correlation
function\cite{denNijs89,Tasaki91,Watanabe93,Nishiyama95,White96_ladder,Kim00}
\begin{equation}\label{eq:strg}
\begin{split}
 O^z_\mathrm{str} (\ell,\ell+2r) := - 
 \Bigg\langle 
  &(S^z_\ell+S^z_{\ell+1})  
\exp\!\left( i\pi \sum_{m=\ell+2}^{\ell+2r-1} S_m^z \right) \\
  &\times (S^z_{\ell+2r}+S^z_{\ell+2r+1})
 \Bigg\rangle.
\end{split}
\end{equation}
The intuition behind this expression is as follows.
Consider a pair of spins $S^z_{\ell+2j}+S^z_{\ell+2j+1}$ on the bond
$(\ell+2j, \ell+2j+1)~(j\in\mathbb{Z})$, which the string correlation
function (\ref{eq:strg}) consists of.
If an odd number of valence bonds cross
any cut placed between the neighboring pairs,
then the pattern
of $ S^z_{\ell+2j}+S^z_{\ell+2j+1}=-1,0,+1$ shows a hidden
antiferromagnetic order, namely, alternation of $+1$ and $-1$ after
removing all $0$'s (see figures in Refs.~\onlinecite{Nishiyama95}
and \onlinecite{Kim00}).
The correlation function \eqref{eq:strg} detects
this hidden order and takes a
non-vanishing value in the long-distance limit $r\to\infty$.
 
%============================
\begin{figure}
\begin{center}
\includegraphics[width=0.48\textwidth]{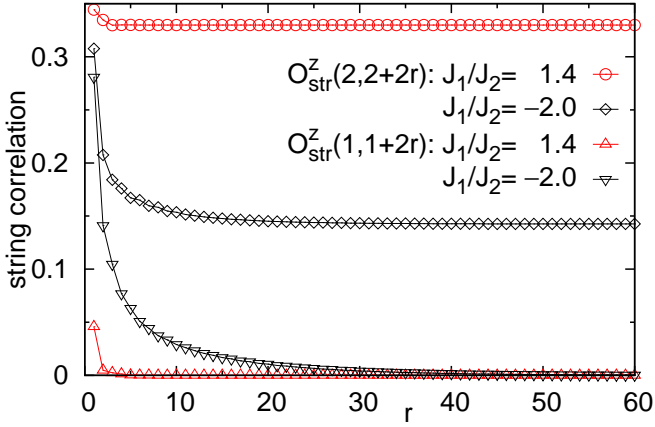}%{strg_z1.eps}
\end{center}
\caption{(Color online) String correlation function \eqref{eq:strg}
for $J_1/J_2=1.4$ and $-2.0$ in the isotropic case $\Delta=1$.  For
both values of $J_1/J_2$, $O^z(2,2+2r)$ remains finite in the
long-distance limit while $O^z(1,1+2r)$ decays to zero.  }
\label{fig:strg_z1}
\end{figure}
%============================

%--------------------------------------
%- Numerical result of string correlation functions

Figure~\ref{fig:strg_z1} presents the numerical data of the string
correlation functions \eqref{eq:strg} calculated with different
starting points $\ell=1,2$
for the ground state with $D_{123}<0$.
We find that for both signs of $J_1$,
$O^z(2,2+2r)$ remains finite in the long-distance limit while
$O^z(1,1+2r)$ decays to zero, in agreement with the even-odd
structure in Fig.~\ref{fig:dimer}.
We note that this behavior is also
consistent with the bosonized expressions of the string
correlations\cite{Nakamura03}
\begin{align}
 &O^z_\mathrm{str} (1,1+2r) \sim
 \langle \cos[\sqp\phi_+(x)] \cos[\sqp\phi_+(y)] \rangle ,\\
 &O^z_\mathrm{str} (2,2+2r) \sim
 \langle \sin[\sqp\phi_+(x)] \sin[\sqp\phi_+(y)] \rangle,
\end{align}
(with $x$ and $y$ being the two endpoints of the string) 
and the field locking position $\sqfp\phi_+=\pi$ for the ground state
with $D_{123}<0$ [see Eq.~\eqref{eq:lock_dimer}]. 
The $J_1/J_2$-dependence of $O^z(2,2+2r)$ for a long distance $r=50$ 
is shown in Fig.~\ref{fig:dim_z1}(c). 
Although the dimer order parameter shows a large difference in magnitude 
between the $J_1>0$ and $J_1<0$ cases, 
the values of the string correlation are rather comparable 
between the two cases.  

%============================
\begin{figure}
\begin{center}
\includegraphics[width=0.48\textwidth]{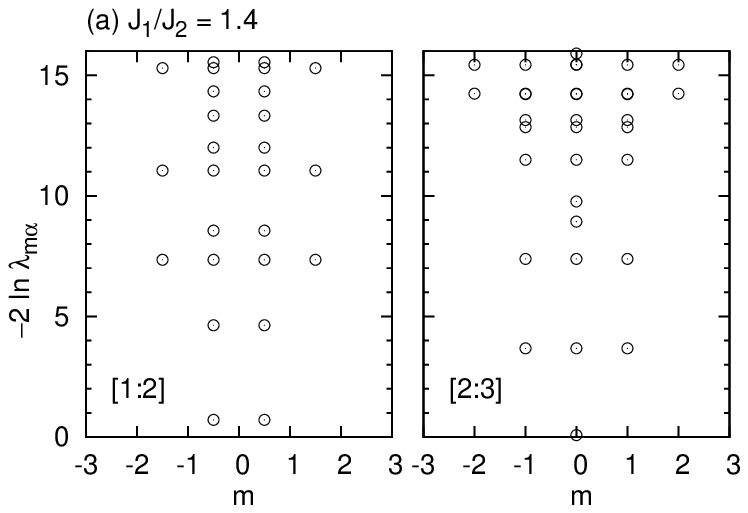}\\%{ES_jp1p4.eps}
\includegraphics[width=0.48\textwidth]{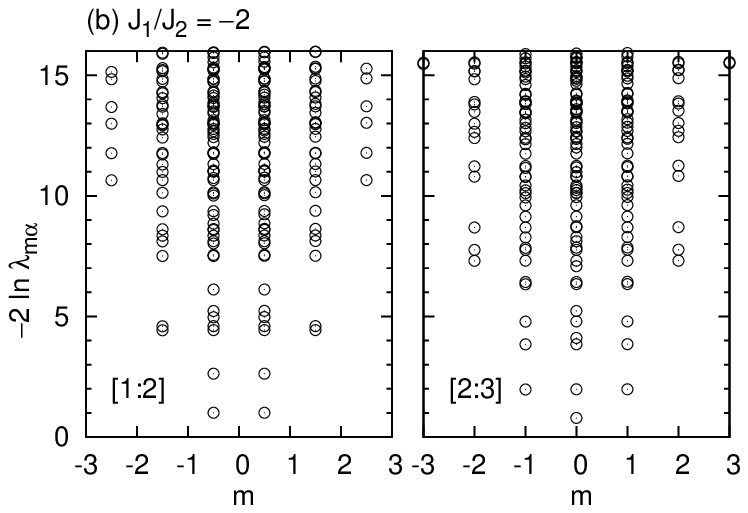}%{ES_jm2p0.eps}
\end{center}
\caption{(Color online) Entanglement spectra $\{-2 \ln
\lambda_{m\alpha}\}$ for (a) $J_1/J_2=1.4$ and (b) $J_1/J_2=-2$.  The
left and right panels are for the bipartition of the system at the
bonds $(1,2)$ and $(2,3)$, respectively.  $m$ refers to the
magnetization in the right half of the system.  The lower entanglement
level corresponds to the more important weight in the total state.  We
note that as an example, the exact singlet dimer ground state of
Ref.~\onlinecite{Majumdar69} shows $-2\ln \lambda_{\pm\frac12,1}=\ln
2$ and $-2\ln \lambda_{0,1}=0$ for the two types of bipartition (with
all the other levels at infinity).  }
\label{fig:ES}
\end{figure}
%============================

%--------------------------------------
%- Entanglement spectrum

Another way of probing the hidden order is to find the degeneracy in
the entanglement spectrum.\cite{Pollmann10} Using the Schmidt
coefficients $\{\lambda_{m\alpha}\}$ calculated in iTEBD, we plot the
spectra $\{-2 \ln \lambda_{m\alpha}\}$ in Fig.~\ref{fig:ES}.  Here
the spectra are classified by the $z$-component magnetization $m$ in
the right half of the system (this classification is done in the
process of our calculations to exploit the U(1) spin rotational
symmetry for better efficiency).  For the bipartition of the system at
the bond $(1,2)$ (left panels), we find that the entanglement levels
appear only for half-integer $m$, and are all doubly degenerate due to
the left-right symmetry around $m=0$.  By contrast, for the
bipartition at $(2,3)$ (right panels), the entanglement levels appear
only for integer $m$, and non-degenerate levels are found for
$m=0$.\cite{Comment_ES} These features are found commonly for both
signs of $J_1$, and are consistent with the even-odd structure in
Fig.~\ref{fig:dimer}.

%--------------------------------------
%- Longer-range valence bonds

In Fig.~\ref{fig:dimer}, we
depicted short-range valence bonds only. 
However, the even-odd structure we discussed can be also 
defined in the presence of longer-range valence bonds.
As the correlation length becomes longer, the weights of such
longer-range valence bonds in the wave function would gradually grow
while retaining the even-odd structure.\cite{Bonesteel89}
We expect that through this process, the Haldane dimer state of
Fig.~\ref{fig:dimer}(b) smoothly changes into the exact
resonating valence bond ground state at $J_1/J_2=-4$,
in which valence bonds are uniformly distributed over
all distances.\cite{Hamada88}

%------------------------------------------------
\subsubsection{Adiabatic connectivity to a ladder model}
%------------------------------------------------

%============================
\begin{figure}
\begin{center}
\includegraphics[width=0.5\textwidth]{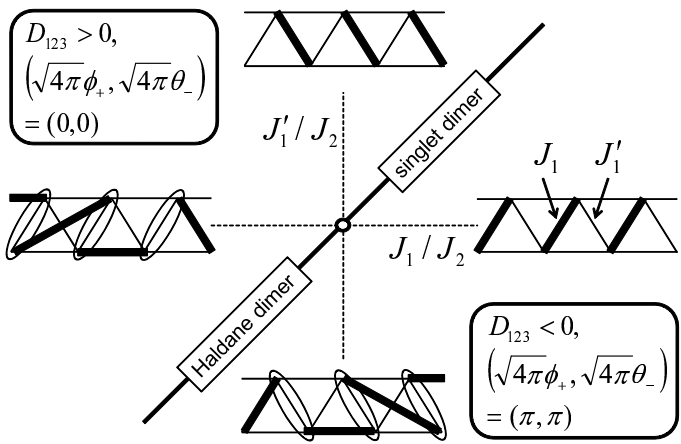}%{J1J1pladder.eps}
\end{center}
\caption{
Expected phase diagram of the zigzag ladder model 
with alternating nearest-neighbor couplings $J_1$ and $J_1'$.  
$|J_1|/J_2$ and $|J_1'|/J_2$ are assumed to be small. 
The solid diagonal line $J_1=J_1'$ corresponds to the original
model \eqref{eq:H} (with $\Delta=1$), 
and represents the first-order phase transition line in the current model. 
The vertical and horizontal dashed lines correspond to a usual ladder model 
(no phase transition on these lines). 
In four insets of zigzag ladders, thick lines indicate valence bonds, 
and ovals indicate the formation of effective spin-$1$'s.
}
\label{fig:J1J1pladder}
\end{figure}
%============================

In order to gain further intuition about the two dimer phases,
it is useful to introduce explicit bond alternation of the $J_1$
couplings in the Hamiltonian \eqref{eq:H} (with $\Delta=1$).
Namely, we place inequivalent couplings $J_1$ and $J_1'$ on the bonds
$(2j+1,2j+2)$ and $(2j+2,2j+3)~(j\in \mathbb{Z})$, respectively.
Figure~\ref{fig:J1J1pladder} displays an expected phase diagram for
small $J_1/J_2$ and $J_1'/J_2$.
This phase diagram can be obtained\cite{Kim00,Kim08} by noticing
that in the non-Abelian bosonization framework,
the bond alternation induces the relevant term
$(J_1-J_1')\Nvec_1\cdot\Nvec_2$ with scaling dimension $1$
in the Hamiltonian, which leads to the ground state where
$D_{123}\sim \langle \Nvec_1\cdot\Nvec_2\rangle$ acquires a finite
average with the same sign as that of $J_1'-J_1$.
The limit $J_1\to 0$ or $J_1'\to 0$ (the vertical or horizontal axis of
Fig.~\ref{fig:J1J1pladder}) corresponds to a spin ladder model, for
which it is established that the rung singlet and
Haldane phases appear for antiferromagnetic and ferromagnetic rung
couplings, respectively.\cite{Dagotto92,Nishiyama95,White96_ladder,Shelton96}
Therefore, we expect that the Haldane dimer state with $D_{123}<0$ in
Fig.~\ref{fig:dimer}(b) should be adiabatically
connected to the Haldane
state of a ladder model (the lower half of the vertical axis of
Fig.~\ref{fig:J1J1pladder}) by gradually switching off the $J_1$
coupling.  It is also possible to adiabatically change
the ground state from the
Haldane dimer state to the singlet dimer state (both with $D_{123}<0$)
by moving counterclockwise around the origin in
Fig.~\ref{fig:J1J1pladder}, although the wave function
may considerably change in this process.
In the zigzag ladder model with $J_1=J_1'$ (diagonal line),
however, the singlet and Haldane dimer phases are
separated by the origin (open circle in Fig.~\ref{fig:J1J1pladder}),
at which the two chains are decoupled.  We note that only on the
$J_1=J_1'$ line in Fig.~\ref{fig:J1J1pladder}, the model has the
symmetry with respect to the translation $\Svec_\ell \to \Svec_{\ell
+1}$, and the dimer order appears by spontaneously breaking this
symmetry.  It would thus be interesting to investigate under what kind
of {\em translationally symmetric} perturbation the Haldane and
singlet dimer phases can be adiabatically connected to each other
while retaining the double degeneracy below a finite excitation gap.

%%%%%%%%%%%%%%%%%%%%%%%%%%%%%%%%%%%%%%%%%%%%%%%%%
\section{Easy-plane case $0\le \Delta<1$} \label{sec:easy-plane}
%%%%%%%%%%%%%%%%%%%%%%%%%%%%%%%%%%%%%%%%%%%%%%%%%

In this section, we consider the model \eqref{eq:H} in the easy-plane
case $0\le\Delta<1$.  In Sec.~\ref{subsec:bos_easy-plane}, we present
the Abelian bosonization formulation of the model for $|J_1|/J_2 \ll
1$ and explain how various phases in Fig.~\ref{fig:phase} are
described in this framework.  In particular, we review the effective
theory for the gapless chiral phase\cite{Nersesyan98} and, following
Ref.~\onlinecite{Lecheminant01}, discuss its instability towards
gapped chiral phases due to a symmetry-allowed perturbation.
Section~\ref{subsec:numerics_easy-plane} presents our numerical
results.  We compute the spin correlation functions in the gapless chiral
phase and determine the phase boundaries
to the gapped chiral phases.

%************************************************
\subsection{Bosonization analyses}\label{subsec:bos_easy-plane}
%************************************************

We consider the easy-plane XXZ Hamiltonian \eqref{eq:H} in the regime
$|J_1|/J_2\ll 1$.  Using the formulation described in
Sec.~\ref{subsec:isotropic_abel_bos}, we obtain the effective
Hamiltonian
\begin{equation}\label{eq:Heff}
\begin{split}
 H = \int dx\Big\{
     & \sum_{\nu=\pm}
       \frac{v_\nu}{2}
       \left[ K_\nu (\partial_x\theta_\nu)^2
             + K_\nu^{-1} (\partial_x\phi_\nu)^2 \right]\\
     &-\gamma_1 \cos(\sqfp\phi_+) \cos(\sqfp\theta_-)\\
     &+\gamma_\tw (\partial_x \theta_+) \sin(\sqfp\theta_-)\\
     &\left.{}+\gamma_\tw'' (\partial_x \phi_-) \sin(\sqfp\phi_+)+ \ldots
      \right\}.   
\end{split}
\end{equation}
The first line represents the Gaussian Hamiltonian while the other
lines represent perturbations which can become relevant
in the easy-plane case.\cite{Nersesyan98,FSSO08,Cabra00}
As seen in Eq.~\eqref{eq:O_bos}, the $\gamma_1$ term is
related to the $G_1$ term in the non-Abelian bosonization,
while $\gamma_\tw$ and $\gamma_\tw''$ correspond to $G_\tw$.
The coupling constants are obtained in lowest order
in $J_1$ as
\begin{align}
 &K_\pm = K \left( 1\mp \frac{K J_1 \Delta a}{2\pi v} \right),
 v_\pm = v \left( 1\pm \frac{K J_1 \Delta a}{2\pi v}  \right),
\label{eq:K and v}
\\ 
 &\gamma_1 = \frac{B_1^2J_1}{a},~~
 \gamma_\tw = \sqp J_1 B_0^2,~~
 \gamma_\tw'' = \frac{\sqp}{2} J_1\Delta A_1^2,
\end{align}
where $K$ and $v$ are given by Eq.~\eqref{eq;K_v_bos}. 
We have discussed in Sec.~\ref{sec:dimer} that,
in the isotropic case $\Delta=1$, $\gamma_1$ grows to large positive
values for both signs of $J_1$ under the RG, 
and induces the singlet and Haldane dimer phases 
for $J_1>0$ and $J_1<0$, respectively. 
Below we explain how other phases in Fig.~\ref{fig:phase} 
are described using the effective Hamiltonian \eqref{eq:Heff}. 
The results are summarized in Table~\ref{table:bos}. 

\newcommand{\nd}{\mathrm{nd}}
\newcommand{\twolines}[2]{$\begin{matrix}\text{#1}\\\text{#2}\end{matrix}$}
%##################
\begin{table*}
\caption{\label{table:bos} Summary of the Abelian bosonization
description of the phases for small $|J_1|/J_2$.  Both the easy-plane
(Sec.~\ref{sec:easy-plane}) and easy-axis (Sec.~\ref{sec:easy-axis})
cases are presented.
%$c_{1,2,3}$ are non-zero constants.
We note that the (chiral) even-parity dimer and chiral N\'eel phases appear
for rather large $|J_1|/J_2$ in Fig.~\ref{fig:phase} although their
essential features can be captured in the Abelian bosonization
framework.  }
\begin{tabular}{cccc}
\hline\hline
 Phase
 & Relevant perturbations  
 & Field-locking positions
 & Order parameters \\ 
\hline
 Singlet/Haldane dimer
 & $\gamma_1>0$  %($\gamma_\tw\ne 0$)
 & $(\sqfp\phi_+,\sqfp\theta_-)=(0,0),~(\pi,\pi)$
 & $D^{xy}_{123}D^z_{123}>0$\\
 Even-parity dimer
 & $\gamma_1<0$  %($\gamma_\tw< 0$)
 & $(\sqfp\phi_+,\sqfp\theta_-)=(0,\pi),~(\pi,0)$
 & $D^{xy}_{123}D^z_{123}<0$\\
 Gapless chiral
 & $\gamma_\tw\sim J_1\ne 0$
 & $\sqfp\theta_-=-\frac{\pi}{2} \mathrm{sgn}
  (J_1\langle\partial_x\theta_+\rangle) $
 & $\kappa_{12}^z\ne 0$\\
 Chiral singlet/Haldane dimer
 & $\gamma_\tw\sim J_1$, $\gamma_\nd<0$, $\gamma_1>0$
 & 
$
(\sqfp\phi_+,\sqfp\theta_-)=
\begin{cases}
 (0,\pm \frac{\pi}2) \to (0,0)\\
 (\pi,\pm \frac{\pi}2) \to (\pi,\pm\pi)
\end{cases}
$
 & $\kappa_{12}^z\ne 0$, $D^{xy}_{123}D^z_{123}>0$\\
 Chiral even-parity dimer
% & \twolines{$\gamma_\tw< 0$,}{$\gamma_\nd<0$, $\gamma_1<0$}
 & $\gamma_\tw< 0$, $\gamma_\nd<0$, $\gamma_1<0$
 & 
$
(\sqfp\phi_+,\sqfp\theta_-)=
\begin{cases}
 (0,\pm \frac{\pi}2) \to (0,\pm\pi)\\
 (\pi,\pm \frac{\pi}2) \to (\pi,0)
\end{cases}
$
 & $\kappa_{12}^z\ne 0$, $D^{xy}_{123}D^z_{123}<0$\\
%\hline
 Chiral N\'eel
 & $\gamma_\tw< 0$, $\gamma_\nd>0$, $\gamma_\tw''\ne 0$
 & $\sqfp\phi_+=\pm \pi/2$, $\sqfp\theta_-=\pm\pi/2$
 & $\kappa_{12}^z\ne 0$, $\langle S_\ell^z \rangle\propto(-1)^\ell$\\
%\hline
 uudd
 & $\gamma_\bs<0$
% & $\sqtp(\phi_1,\phi_2)=(0,0),(0,\pi),(\pi,0),(\pi,\pi)$
 & $\sqtp\phi_1=0,\pi$, $\sqtp\phi_2=0,\pi$
 & $\langle S_{2j+1}^z\rangle=\pm\langle S_{2j}^z\rangle
   \propto(-1)^j$\\
% & \twolines{$\langle S_{2j+1}^z \rangle=
% c_2(-1)^j$,}{$\langle S_{2j+2}^z \rangle=\pm c_2(-1)^j$} \\
%hline
 Partially polarized
 & $\gamma_\tw'<0$
 & $\sqfp\phi_-=\frac{\pi}2 \mathrm{sgn}(\langle \partial_x\phi_+\rangle)$
 & $\langle S_\ell^z \rangle\ne0$\\
\hline\hline
\end{tabular}
\end{table*}
%##################

%------------------------------------------------
\subsubsection{Even-parity dimer phase}
%------------------------------------------------

If $J_1<0$, 
the coupling constant $\gamma_1$ is negative at the bare level. 
Suppose that this term grows, keeping the negative sign under the RG.  
Then the bosonic fields are locked at 
\begin{equation}\label{eq:lock_tridimer}
 (\sqfp\phi_+,\sqfp\theta_-)=(0,\pi) ~~\text{or}~~ (\pi,0). 
\end{equation}
In either case, it follows from Eq.~\eqref{eq:bos_dimer}
that the $xy$ and $z$ components of the dimer order parameter,
$D_{123}^{xy}$ and $D_{123}^z$, become finite
and have mutually opposite signs ($D_{123}^{xy} D_{123}^z<0$).
This situation corresponds to the even-parity dimer phase appearing 
at strong easy-plane anisotropy ($\Delta\lesssim 0.6$); 
see Fig.~\ref{fig:order_jm2}(b). 

%------------------------------------------------
\subsubsection{Gapless chiral phase}\label{sec:bos_gapless_chiral}
%------------------------------------------------

%--------------------------------------
% Mean-field theory a la Nersesyan et al.

As shown by Nersesyan {\em et al.},\cite{Nersesyan98} 
the gapless chiral phase appears when $\gamma_\tw$ grows under the RG. 
To discuss the effect of the $\gamma_\tw$ term,
it is useful to perform the mean-field decoupling\cite{Nersesyan98}
\begin{equation}
\begin{split}
 &(\partial_x\theta_+) \sin(\sqfp \theta_-) \\
 &\to
 \langle \partial_x\theta_+ \rangle \sin(\sqfp \theta_-)
 +(\partial_x\theta_+) \langle \sin(\sqfp \theta_-) \rangle.
\end{split}
\end{equation}
Then the Hamiltonian \eqref{eq:Heff} separates into ``$+$'' and
``$-$'' sectors: 
\begin{equation}
 H=H_+ + H_-
\end{equation} 
with
\begin{align}
 H_+ &= \int dx \frac{v_+}2
          \left[ K_+ (\partial_x\thetat_+)^2
               + K_+^{-1} (\partial_x\phi_+)^2 \right],\\
 H_- &= \int dx \Big\{ \frac{v_-}2
          \left[ K_- (\partial_x\theta_- )^2
               + K_-^{-1} (\partial_x\phi_-)^2 \right] \notag \\
  & \qquad\qquad
   + \gamma_\tw \langle \partial_x \theta_+ \rangle \sin (\sqfp \theta_-)
    \Big\}. \label{eq:chiral_H-} 
\end{align}
Here we have introduced
\begin{equation} \label{eq:thetat}
 \thetat_+ := \theta_+ -q x, \qquad
 q:=  - \frac{\gamma_\tw \langle \sin (\sqfp \theta_-) \rangle}{v_+ K_+}.
\end{equation}
While $H_+$ is a Gaussian Hamiltonian of free bosons
$(\phi_+,\thetat_+)$,
$H_-$ is a sine-Gordon Hamiltonian in which the relevant sine potential
generates a finite energy gap for the $\theta_-$ field.
Since $\langle \partial_x \thetat_+ \rangle =0$ from $H_+$, 
$\langle \partial_x \theta_+ \rangle = q$. 
The coefficient of the sine potential in $H_-$ is 
thus given by $\gamma_\tw q$, 
and the field $\theta_-$ is locked at distinct positions
depending on the sign of this coefficient: 
\begin{equation} \label{eq:thetam_chiral}
  \langle \sqfp\theta_- \rangle = -\frac{\pi}2 {\rm sgn}( \gamma_\tw q ).
\end{equation}
Correspondingly, the sine term acquires a finite expectation value:
\begin{equation}\label{eq:sin_thetam}
 \langle \sin (\sqfp \theta_-) \rangle = -c_1 ~{\rm sgn}( \gamma_\tw q ),
\end{equation}
where $c_1$ is a positive constant. 
Equations~\eqref{eq:thetat} and \eqref{eq:sin_thetam} can be
solved self-consistently\cite{Kolezhuk05} 
by inserting the exact solution of the sine-Gordon model into Eq.~\eqref{eq:sin_thetam},
yielding two solutions,
one positive and one negative $q$. 
It should be understood that the mean-field parameters $c_1$ and $q$
used in the following calculation of correlation functions are
determined selfconsistenly.

%--------------------------------------
% Physical properties of the gapless chiral phase

First, the non-vanishing value of the mean-field parameter 
in Eq.~\eqref{eq:sin_thetam} directly leads to
a finite vector chiral order parameter \eqref{eq:chirality}:
\begin{align}\label{eq:kappa_chiral}
 & \kappa^z_{\ell,\ell+1} = -B_0^2  \langle \sin (\sqfp \theta_-) \rangle 
= B_0^2 c_1 ~{\rm sgn}( \gamma_\tw q ). 
% &\langle \kappa_\ell^{(2)} \rangle = - \sqrt{\frac{\pi}{2}} c_2 \langle \partial_x \theta_+ \rangle = - \sqrt{\frac{\pi}{2}} c_2 q.
\end{align}
Therefore the two mean-field solutions correspond to the ground states
with positive and negative $\kappa_{\ell,\ell+1}^z$. 
Let us take the ground state with $\kappa_{\ell,\ell+1}^z>0$ 
(i.e., $\gamma_\tw q>0$) and discuss the expressions of the spin operators. 
We focus on gapless degrees of freedom, and ignore the fluctuations of 
$\theta_-$ around its average \eqref{eq:thetam_chiral}. 
Then we find
\begin{align}
 \sqtp\theta_{1,2} = \sqp (\theta_+ \pm \theta_-) = \sqp\thetat_+ 
+ \sqp q x_{1,2} \mp \frac\pi4, 
\end{align}
which are combined into
\begin{equation}
 \sqtp\theta_n(x_n) = \sqp\thetat_+(x_n) + \sqp q x_n
                      + \frac\pi2 \left( n-\frac32 \right).   
\end{equation}
The in-plane component of the spins are then expressed as
\begin{equation}
\begin{split}
 S_{2j+n}^+ 
 &\approx B_0 (-1)^j e^{i\sqtp \theta_n(x_n)} \\
 &= B_0 \exp \!\left\{ i \!\left[ \sqp (\thetat_+ + q x_n)
             + \frac\pi2 \!\left( 2j+n-\frac32 \right) \right] \right\}. 
\end{split}
\end{equation}
Introducing $\ell =2j+n$ and $x(\ell)=x_n(j)=(a/2) (\ell-3/2)$, we obtain
\begin{equation}
 S_\ell^+ \approx B_0 e^{i [\sqp\thetat_+(x) + Q(\ell-3/2)]},  
\end{equation} 
with 
\begin{equation}
 Q = \frac{\pi + \sqp q a}2. 
\end{equation}
As for the $z$ component of the spins,
we simply ignore the $\phi_-$ part of the expression:
\begin{equation}
 S_\ell^z \approx \frac{a}{\sqrt{4\pi}} \partial_x \phi_+. 
\end{equation}
Spin correlation functions are then calculated as\cite{Nersesyan98,Hikihara08}
\begin{align}
 &\langle S^+_\ell S^-_{\ell'} \rangle
   = A \frac{e^{-iQ(\ell'-\ell)}}{|\ell'-\ell|^{1/(2K_+)}}
     +\dots, \label{eq:Spm_chiral} \\
 &\langle S^z_\ell S^z_{\ell'} \rangle
   = - \frac{K_+}{2\pi^2 |\ell'-\ell|^2} + \dots~\label{eq:Szz_chiral}
\end{align}
with $A=B_0^2 2^{1/(2K_+)}$. 
The finite vector chiral order parameter $\kappa^z_{\ell,\ell+1}$
in Eq.~\eqref{eq:kappa_chiral} 
and the quasi-long-range in-plane spiral correlation
with an incommensurate pitch angle $Q$ in Eq.~\eqref{eq:Spm_chiral}
are two major features of the gapless chiral phase.

%------------------------------------------------
\subsubsection{Gapped chiral phases} \label{sec:bos_gap_ch}
%------------------------------------------------

%--------------------------------------
% Introduction of vertex term a la Lecheminant et al.

Following Lecheminant {\it et al.},\cite{Lecheminant01}
we consider the following symmetry-allowed perturbation to
the effective theory of the gapless chiral phase: 
\begin{equation}
 \gamma_\nd \int dx \cos (2\sqfp \phi_+),
\end{equation}
with which 
the ``+'' sector of the Hamiltonian becomes a sine-Gordon model.
The scaling dimension of this perturbation is $4K_+$.
If the $\gamma_\nd$ term becomes relevant ($4K_+<2$), 
a Berezinskii-Kosterlitz-Thouless (BKT) transition takes place 
and as a result, the bosonic field $\phi_+$ is locked at distinct
positions dependent on the sign of $\gamma_\nd$.
This leads to gapped chiral phases in which the
chiral order coexist with either the dimer or the N\'eel order,
depending on the sign of $\gamma_\nd$.

%--------------------------------------
% Chiral dimer phases

First, when $\gamma_\nd<0$, $\sqfp \phi_+$ is locked at
\begin{equation} \label{eq:phip_chiraldimer}
 \sqfp \phi_+ = 0~\text{or}~\pi,  
\end{equation} 
which produces a finite value of the $z$-component of
the dimer order parameter, $D^z_{123}$, 
as seen in Eq.~\eqref{eq:bos_dimer}. 
We have thus obtained the ``chiral dimer phase,''
in which the vector chiral and dimer orders coexist.\cite{Lecheminant01} 
Once $\phi_+$ is locked as in Eq.~\eqref{eq:phip_chiraldimer}, 
the locking position of $\sqfp \theta_-$ is affected
by the $\gamma_1$ term in Eq.~\eqref{eq:Heff} and changed from
$\pm \pi/2$ of the gapless chiral phase [Eq.~\eqref{eq:thetam_chiral}],
so that the $xy$-component of the dimer order parameter, $D^{xy}_{123}$, 
also becomes finite,
in agreement with Fig.~\ref{fig:order_jm2}. 
Specifically, for positive $\gamma_1$, the field-locking positions of  
the four degenerate ground states change smoothly
with the strength of $\gamma_1$ as
\begin{equation}
(\sqfp\phi_+,\sqfp\theta_-)=
\begin{cases}
 (0,\pm \frac{\pi}2) \longrightarrow (0,0),\\
 (\pi,\pm \frac{\pi}2) \longrightarrow (\pi,\pm\pi) \equiv (\pi,\pi), 
\end{cases}
\end{equation}
finally resulting in the two degenerate ground states
of either the Haldane dimer or the singlet dimer phase
as specified by Eq.~\eqref{eq:lock_dimer}. 
For negative $\gamma_1$, the field-locking positions change as
\begin{equation}
(\sqfp\phi_+,\sqfp\theta_-)=
\begin{cases}
 (0,\pm \frac{\pi}2) \longrightarrow (0,\pm \pi) \equiv (0,\pi) \\
 (\pi,\pm \frac{\pi}2) \longrightarrow (\pi,0), 
\end{cases}
\end{equation}
resulting in the two degenerate ground states of
the even-parity dimer phase as indicated by
Eq.~\eqref{eq:lock_tridimer}. 

%--------------------------------------
% Chiral Neel phase

Second, when $\gamma_\nd>0$, $\phi_+$ is locked at
\begin{equation}\label{eq:lock_chNeel}
 \sqfp \phi_+ = \pm \frac\pi2.   
\end{equation} 
This yields a finite N\'eel order paramter along the $z$ direction,
as we explain below.
From Eqs.~\eqref{eq:Sz_bos} and \eqref{eq:phi_theta_pm},
the $S^z_\ell$ operator has the staggered component
\begin{equation} \label{eq:Sz_stagger}
 (-1)^\ell S^z_\ell = \frac{a}{\sqrt{4\pi}} \partial_x \phi_- + \dots,
\end{equation}
which, at first sight,
looks insensitive to the locking of $\phi_+$. 
However, after the locking \eqref{eq:lock_chNeel}, 
the $\gamma''_\tw$ term in Eq.~\eqref{eq:Heff}
reduces to the operator $\pm \gamma''_\tw\partial_x\phi_-$, 
which can be absorbed into the Gaussian part of $H_-$ in 
Eq.~\eqref{eq:chiral_H-} by redefining $\phi_-$ 
(so that $\partial_x \phi_-$ is shifted by a constant). 
Consequently, Eq.~\eqref{eq:Sz_stagger} acquires a
nonvanishing
expectation value 
\begin{equation}
  (-1)^\ell \langle S^z_\ell \rangle =
    - \frac{K_+ a}{\sqfp v_+} \gamma''_\tw \langle\sin(\sqfp\phi_+)\rangle
    +\dots .
\end{equation}
We have therefore obtained the ``chiral N\'eel phase,''
in which the vector chiral and N\'eel orders coexist. 

%--------------------------------------
% Logarithmic corrections at the KT transition

At the BKT transition point $K_+=1/2$, 
the sine-Gordon theory for the ``+'' sector predicts
the appearance of 
a multiplicative logarithmic correction to the correlation 
functions:\cite{Giamarchi04,Kosterlitz74,Giamarchi88,Eggert96}
\begin{align}\label{eq:Spm_bos_log}
 \langle S^+_\ell S^-_{\ell'} \rangle =
   A \frac{e^{-iQ(\ell'-\ell)}}{|\ell'-\ell|} \ln^{1/2}( |\ell'-\ell|/a )
    +\dots~.
\end{align}
This logarithmic correction is utilized to locate the BKT
phase transition point numerically in the next section. 

%============================
\begin{figure}
\begin{center}
\includegraphics[width=0.48\textwidth]{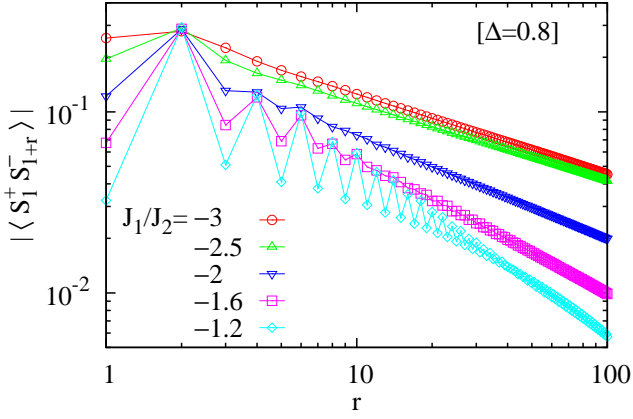}%{Cpm_z0p8.eps}
\end{center}
\caption{(Color online) In-plane spin correlation function $|\langle
S^+_1 S^-_{1+r}\rangle|$ for fixed $\Delta=0.8$ and various values of
$J_1/J_2$ in the gapless chiral phase.  Logarithmic scales are used in
both axes.  }
\label{fig:Cpm_z0p8}
\end{figure}
%============================

%============================
\begin{figure}
\begin{center}
\includegraphics[width=0.48\textwidth]{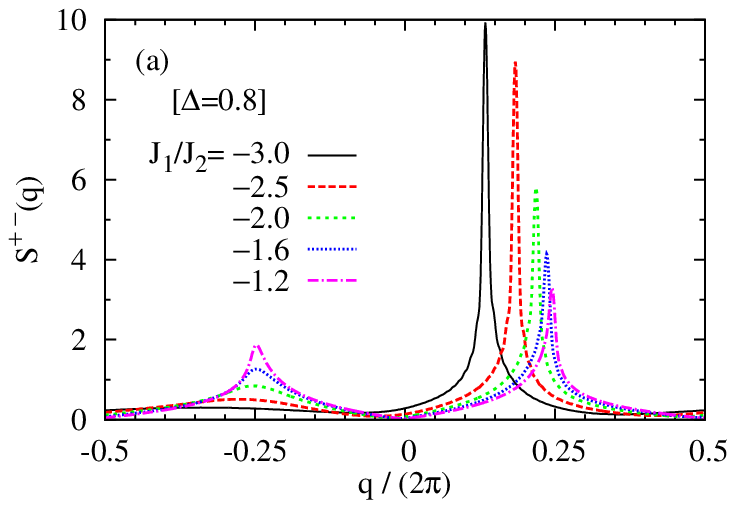}\\%{Spm_z0p8.eps}
\includegraphics[width=0.48\textwidth]{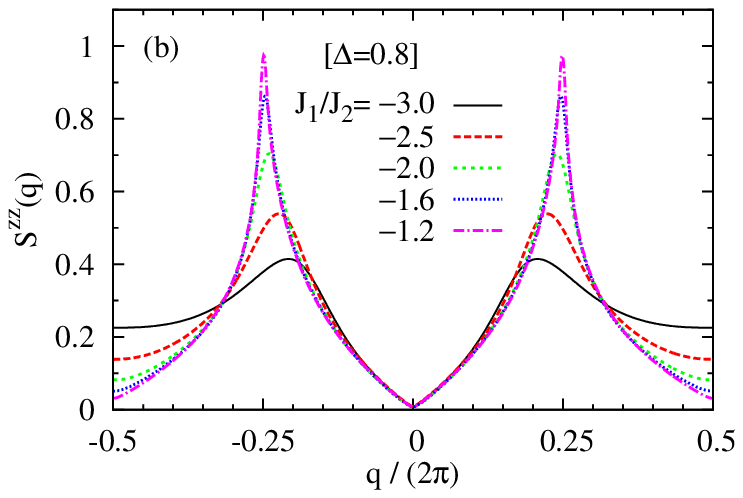}%{Szz_z0p8.eps}
\end{center}
\caption{(Color online) Equal-time spin structure factors
[Eq.~\eqref{eq:Sq_def}], (a) $S^{+-}(q)$ and (b) $S^{zz}(q)$, in the
gapless chiral phase.  Calculations were done for the same parameter
points as in Fig.~\ref{fig:Cpm_z0p8}, and we set $L=100$.  }
\label{fig:Spm_z0p8}
\end{figure}
%============================

%************************************************
\subsection{Numerical results}\label{subsec:numerics_easy-plane}
%************************************************

%============================
\begin{figure}
\begin{center}
\includegraphics[width=0.48\textwidth]{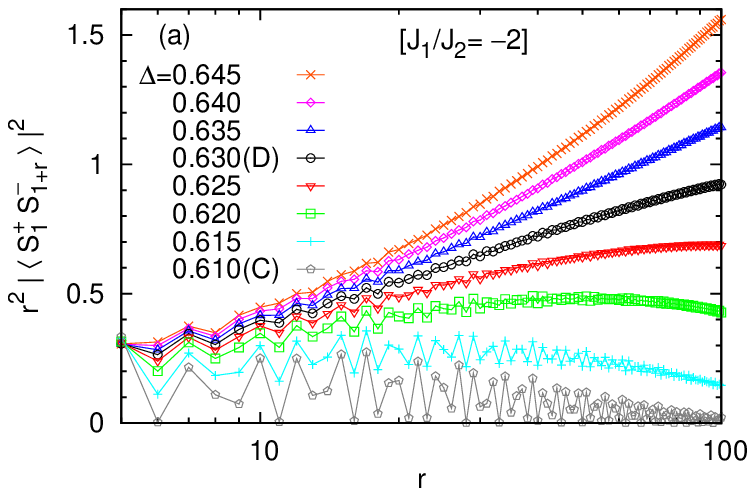}\\%{Cpm_z0p6jm2.eps}
\includegraphics[width=0.48\textwidth]{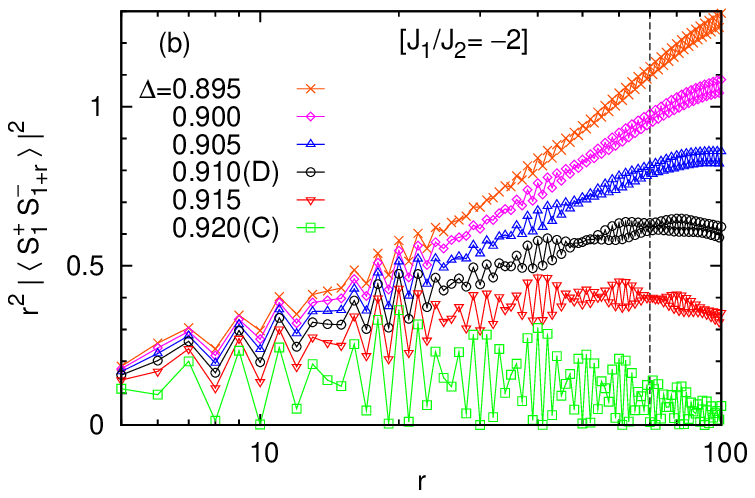}%{Cpm_z0p9jm2.eps}
\end{center}
\caption{(Color online) In-plane spin correlation function, calculated
for fixed $J_1/J_2=-2$ and various values of $\Delta$ around the
transition points shown in Fig.~\ref{fig:order_jm2}.  The symbols
``C'' and ``D'' indicate our estimates of the transition points (with
a precision of 0.005) for the onsets of the vector chiral and dimer
orders, respectively.  Logarithmic scale is used for the horizontal
axis.  At the BKT transition related to the onset of the dimer order,
the plotted function is expected to become linear in the long-distance
limit, which we use to determine the ``D'' points.  In panel (b), all
the curves are slightly bent downward around $r=70$ (broken vertical
line) due to the finiteness of the Schmidt rank $\chi(=300)$ in iTEBD,
so we use the range $r\lesssim 70$ for our analysis.  }
\label{fig:Cpm_jm2}
\end{figure}
%============================

%============================
\begin{figure}
\begin{center}
\includegraphics[width=0.48\textwidth]{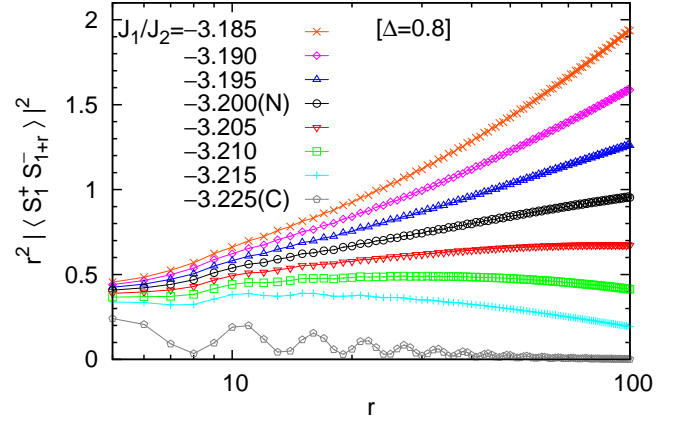}%{Cpm_z0p8jm3.eps}
\end{center}
\caption{(Color online) In-plane spin correlation, calculated for
fixed $\Delta=0.8$ and various values of $J_1/J_2$ around the
transition points shown in Fig.~\ref{fig:order_z0p8}.  The symbols
``C'' and ``N'' indicate our estimates of the transition points (with
a precision of 0.005) for the onsets of the vector chiral and N\'eel
orders, respectively.  }
\label{fig:Cpm_z0p8jm3}
\end{figure}
%============================

%--------------------------------------
%- Subsection introduction

In this section, we present our numerical iTEBD results (with the
Schmidt rank $\chi=300$) on the spin correlation functions in the
easy-plane case $0\le \Delta <1$.

%------------------------------------------------
\subsubsection{Spin correlations in the gapless chiral phase} 
%------------------------------------------------

We first discuss the numerical results for the gapless
chiral phase,
where we choose the ground state with $\kappa_{12}>0$.
Figure~\ref{fig:Cpm_z0p8} shows
the in-plane spin correlation function $|\langle S^+_1
S^-_{1+r}\rangle|$ at $\Delta=0.8$ for
various values of $J_1/J_2$ in the gapless chiral phase.
The data for $|J_1|/J_2\gtrsim 2$ follow straight lines
in logarithmic scales, in
agreement with the power-law behavior in Eq.~\eqref{eq:Spm_chiral}.
By contrast, the data for $J_1/J_2=-1.6$ and $-1.2$ show some
oscillations at short distances although the overall
behaviors are linear as expected from Eq.~\eqref{eq:Spm_chiral}
(we suspect that the downward bending at large $r$
for $J_1/J_2=-1.2$ is due to a finite Schmidt rank $\chi=300$,
and is not a genuine behavior).

The origin of the oscillations can be found in the spin structure factors
shown in Fig.~\ref{fig:Spm_z0p8}. 
For $L$ consecutive spins at the sites $\ell=1,2,\dots,L$
in a translationally invariant infinite system treated by iTEBD, 
we introduce
\begin{equation}
 S^\alpha_q = \frac{1}{\sqrt{L}} \sum_{\ell=1}^L S^\alpha_\ell e^{-iq\ell}, 
\end{equation}
and define the equal-time spin structure factors as
\begin{equation}\label{eq:Sq_def}
 S^{\alpha\beta} (q) = \langle S^\alpha_q S^\beta_{-q} \rangle~~
 \text{with}~(\alpha,\beta)=(+,-),~(z,z). 
\end{equation}
In Fig.~\ref{fig:Spm_z0p8}(a), $S^{+-}(q)$ shows sharp peaks
at incommensurate wave number $q=Q>0$,
which become sharper and higher for large $|J_1|/J_2$.
This feature is consistent with Eq.~\eqref{eq:Spm_chiral},
provided that $K_+$ becomes larger with increasing $|J_1|/J_2$;
see Eq.~\eqref{eq:K and v}.
These peaks are expected to diverge as $L\to\infty$ in the gapless
chiral phase. 
For small $|J_1|/J_2$, a second peak around $q=-Q<0$ develops, 
which indicates the ellipticity of the spiral correlations 
and is the origin of the oscillating behavior in Fig.~\ref{fig:Spm_z0p8}. 
The appearance of the second peak can be understood
by observing that $S^{+-}(q)$ should gradually become
left-right symmetric as the vector chiral order parameter
$\kappa^z$ decreases.\cite{FSSO08} 
In Fig.~\ref{fig:Spm_z0p8}(b), $S^{zz}(q)$ shows linear behaviors
around $q=0$ as expected from the Fourier transform of
Eq.~\eqref{eq:Szz_chiral}:  $S^{zz}(q)=K_+|q|/2\pi$ for $|q|\ll 1$. 
In addition, it shows finite peaks at incommensurate $q$. 
Although the explanation of these peaks is beyond the scope of
the effective theory, 
their occurrence is rather natural for $\Delta=0.8$, 
since the $xy$ and $z$ components should show similar behaviors 
as the system approaches the isotropic limit $\Delta=1$. 

%Possible interpretations of anomalous spin correlations observed in LiCu$_2$O$_2$\cite{Seki08} 
%in terms of the ellipticity of the spiral correlations have been presented in Refs.~\onlinecite{FSSO08,Katsura08}. 

%------------------------------------------------
\subsubsection{Transitions to the gapped chiral phases} \label{subsec:gapped_chiral}
%------------------------------------------------

Next we analyze how the spin correlation changes at
the transition from the gapless chiral phase 
to the gapped dimer or N\'eel phase. 
The existence of the intermediate gapped chiral phases where
two kinds of orders coexist is anticipated from
the analyses of the order parameters and entanglement entropy
in Figs.~\ref{fig:order_jm2} and \ref{fig:order_z0p8} 
and from the bosonization analysis
of Sec.~\ref{sec:bos_gap_ch}. 
The in-plane spin correlation function is expected to show a multiplicative
logarithmic correction in Eq.~\eqref{eq:Spm_bos_log} at the BKT transition
point from the gapless to gapped chiral phases. 
Therefore, in Figs.~\ref{fig:Cpm_jm2} and \ref{fig:Cpm_z0p8jm3}, 
we plot $r^2 |\langle S^+_1 S^-_{1+r}\rangle|^2$, which is expected
to become a linear function of $\ln r$
at the BKT transition point. 
In Figs.~\ref{fig:Cpm_jm2}(a) and (b),
the symbols ``C'' indicate the Ising transition points 
(determined in Fig.~\ref{fig:order_jm2}) at which
the inversion symmetry is spontaneously broken and
the vector chiral order appears. 
Finding the linear behavior of the plotted functions,
we determine the BKT transition points as indicated by the symbols ``D''. 
Narrow but finite ranges of intermediate phases between ``C'' and ``D''
are found in the intervals $0.61\lesssim \Delta \lesssim 0.63$ and
$0.91 \lesssim \Delta \lesssim 0.92$, 
which we identify with
the ``chiral (even-parity and Haldane) dimer phases.''  
Similarly, we determine the range of the ``chiral N\'eel phase''
in Fig.~\ref{fig:Cpm_z0p8jm3}. 
In this way, we have determined the ``$\times$'' symbols
in Fig.~\ref{fig:phase}.
Since the method of determining the BKT point from
the logarithmic correction to spin correlation, as employed here,
has not been discussed in literature (as far as we know), 
we demonstrate its validity using a simpler example
in Appendix~\ref{app:TLL-dimer}. 

%============================
\begin{figure}
\begin{center}
 \includegraphics[width=0.48\textwidth]{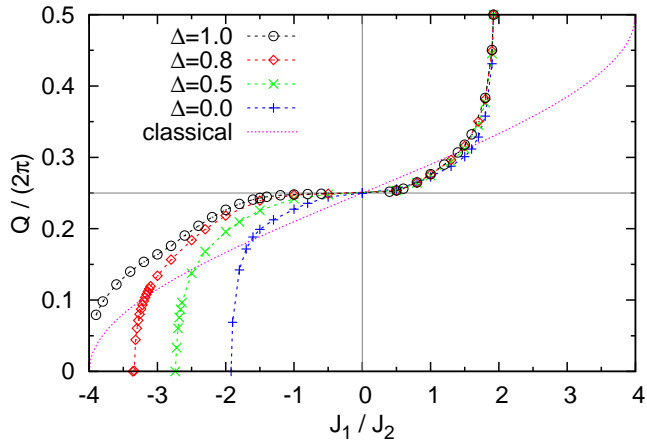}%{pitch_j.eps}
\end{center}
\caption{(Color online) Pitch angle $Q$ as a function of $J_1/J_2$ for
different values of $\Delta$.  This angle is determined by finding the
peak in the in-plane structure factor $S^{+-}(q)$ as shown in
Fig.~\ref{fig:Spm_z0p8}(a).
The classical value $Q=\arccos(-\frac{J_1}{4J_2})$,
which is independent of $\Delta$, is plotted
together for comparison.  }
\label{fig:pitch}
\end{figure}
%============================

%------------------------------------------------
\subsubsection{Pitch angle} \label{subsec:pitch}
%------------------------------------------------

Finally, we determine the pitch angle $Q$ of the incommensurate spin
correlations in the vector chiral and gapped phases. 
It is determined from the maximum position of the in-plane structure
factor $S^{+-}(q)$ (as in Ref.~\onlinecite{Bursill95}). 
The data of $Q$ so obtained as a function of $J_1/J_2$
are shown for different values of $\Delta$ in Fig.~\ref{fig:pitch}. 
The Lifshitz points, at which the in-plane spin correlation function
changes its character from incommensurate to
commensurate ($Q=0$ or $\pi$),   
occur inside the singlet dimer phase for $J_1>0$ 
and inside the even-parity dimer or N\'eel phase for $J_1<0$. 
For $J_1>0$ and all values of $\Delta$, the determined Lifshitz points
are very close to the point $J_1/J_2=2$ with the exact singlet dimer
ground states. 
According to the argument of Ref.~\onlinecite{Nomura05}, 
the Lifshitz points should be in fact located exactly
at $J_1/J_2=2$. 
The small discrepancy comes from the difference in the definition of $Q$; 
in Ref.~\onlinecite{Nomura05}, it is defined in terms of the asymptotic
behavior of the correlation function 
in the long-distance limit. 
For $J_1<0$, the determined Lifshitz line is drawn by broken lines
in Fig.~\ref{fig:phase}; 
it starts from the highly degenerate point\cite{Hamada88,Bursill95}
$(J_1/J_2,\Delta)=(-4,1)$ 
and ends near the point $(J_1/J_2,\Delta)=(-2,0)$
with the exact even-parity dimer ground states.

%%%%%%%%%%%%%%%%%%%%%%%%%%%%%%%%%%%%%%%%%%%%%%%%%
\section{Easy-axis case $\Delta>1$} \label{sec:easy-axis}
%%%%%%%%%%%%%%%%%%%%%%%%%%%%%%%%%%%%%%%%%%%%%%%%%

%--------------------------------------
%- Known numerical results

To complete our analysis of the XXZ chain model \eqref{eq:H} with
$J_1<0$ and $J_2>0$, let us shortly discuss the case with easy-axis
anisotropy $\Delta>1$.  In this case, Igarashi\cite{Igarashi89_FM} and
Tonegawa {\it et al.}\cite{Tonegawa90} have found the following three
phases.  For $J_1/J_2\lesssim -4$, the ground state is fully polarized
(ferromagnetic) along the $z$ direction.  For small $|J_1|/J_2$ and
large $\Delta$, the ground state is antiferromagnetic, having a
period-4 structure $\up\up\down\down\dots$ (uudd).  Between the fully
polarized and uudd phases intervenes the partially polarized phase, in
which the spontaneous ferromagnetic moment along the $z$ direction
changes continuously as a function of $J_1/J_2$ and $\Delta$.  We note
that the uudd phase was also found in the model with antiferromagnetic
$J_{1,2}>0$.\cite{Igarashi89_AFM} Here we describe the uudd and
partially polarized phases in terms of the Abelian bosonization
formulation for $|J_1|/J_2\ll 1$ and $0<\Delta-1 \ll
1$.

%------------------------------------------------
\subsection{uudd phase}
%------------------------------------------------

We start from the decoupled isotropic Heisenberg chains with $J_2>0$.  
The in-chain easy-axis anisotropy
$J_2 (\Delta-1)\sum_{j,n} S^z_{2j+n} S^z_{2j+n+2}$ (with $\Delta>1$) 
adds to the Hamiltonian the backscattering terms 
\begin{equation}
 \gamma_\bs \left[ \cos (2\sqtp\phi_1) + \cos (2\sqtp\phi_2) \right]  
\end{equation}
with $\gamma_\bs<0$. 
If this term grows dominantly under the RG, 
the fields are locked at
\begin{equation}
 (\sqtp\phi_1,\sqtp\phi_2)=(0,0),~(0,\pi),~(\pi,0),~\text{or}~(\pi,\pi).
\end{equation}
These four-fold degenerate ground states 
correspond to the period-4 uudd structures with
\begin{equation}
 \langle S_{2j+1}^z \rangle=c_2(-1)^j,
~\langle S_{2j+2}^z \rangle=\pm c_2(-1)^j,
\end{equation}
where $c_2$ is a non-zero constant [see Eq.~\eqref{eq:Sz_bos}]. 

%------------------------------------------------
\subsection{Partially polarized phase}
%------------------------------------------------

The partially polarized phase found
numerically\cite{Tonegawa90} can be understood from the mean-field
treatment of the operator
$(\partial_x \phi_+) \sin(\sqfp\phi_-)$,\cite{Zarea04}  
which is contained in $\Ocal_\tw$ in Eq.~\eqref{eq:Otw_bos}. 
Here we review the formulation of Zarea {\it et al.},\cite{Zarea04}
and then discuss the behaviors of correlation functions,  
which were not discussed in detail in previous
studies.\cite{Tonegawa90,Zarea04,Igarashi89_FM} 

We start from the effective Hamiltonian\cite{Zarea04}
\begin{equation}\label{eq:Heff_easyaxis}
\begin{split}
 H = \int dx &\Big\{ \sum_{\nu=\pm} \frac{v_\nu}2
             \left[ K_\nu (\partial_x\theta_\nu)^2
                  + K_\nu^{-1} (\partial_x\phi_\nu)^2 \right]\\
     &+\gamma_\tw' (\partial_x \phi_+) \sin(\sqfp\phi_-)\Big\}
\end{split}
\end{equation}
with $\gamma_\tw'<0$. 
The mean-field decoupling similar to the one used
in Sec.~\ref{sec:bos_gapless_chiral}
yields the effective Hamiltonian $H=H_++H_-$, where
\newcommand{\phit}{\tilde{\phi}}
\begin{align}
 H_+ &= \int dx \frac{v_+}2 \left[ K_+ (\partial_x\theta_+)^2
       + K_+^{-1} (\partial_x\phit_+)^2 \right],\\
 H_- &= \int dx \Big\{ \frac{v_-}2
        \left[ K_- (\partial_x\theta_- )^2
             + K_-^{-1} (\partial_x\phi_-)^2 \right] \notag \\
 & \qquad\qquad + \gamma_\tw' \mu \sin (\sqfp \phi_-)\Big\}. 
\end{align}
Here we have introduced
\begin{equation}
 \phit_+(x)=\phi_+(x)-\mu x,
\end{equation}
with
\begin{equation}
 \mu=-\frac{K_+\gamma_\tw' \langle \sin(\sqfp\phi_-) \rangle}{v_+}
    =\langle \partial_x\phi_+\rangle. 
\end{equation}
There are two self-consistent solutions:
$\mu=+|\mu|, -|\mu|$.
A non-vanishing $\mu$ directly leads to the spontaneous magnetization
\begin{equation}
 \langle S_\ell^z \rangle = \frac{a}{\sqfp}\langle \partial_x\phi_+\rangle
  = \frac{a}{\sqfp} \mu \equiv M. 
\end{equation}
Furthermore, the sine potential in $H_-$ locks the bosonic field at
\begin{equation}\label{eq:lock_pFM}
 \sqfp\phi_- = \frac{\pi}2 {\rm sgn}(\mu). 
\end{equation}

To see the physical consequence of the field locking in
Eq.~\eqref{eq:lock_pFM}, we discuss spin correlation functions in the
ground state.  The transverse component of spin,
$S_\ell^+$, contains the operator $e^{\pm\sqp\theta_-}$,
which strongly fluctuates due to the locking
of the dual field $\phi_-$;
therefore the correlation function
$\langle S_\ell^+ S_{\ell'}^- \rangle$ decays exponentially
with the distance.
Instead, the longitudinal correlation
$\langle S_\ell^z S_{\ell'}^z \rangle$ and the bond nematic
correlation\cite{Hikihara08,Sudan09}
$\langle S_\ell^+ S_{\ell+1}^+ S_{\ell'}^- S_{\ell'+1}^- \rangle$
show power-law decays.
Ignoring fluctuations of $\phi_-$,
we obtain the bosonized expressions for 
these operators as
\begin{align}
 &S_\ell^z = M + \frac{a}{\sqfp} \phit_+ \notag \\
 &\qquad
   + A_1 \cos \!\left[ \sqp \phit_+
                   + \pi\!\left(M- \frac{\mathrm{sgn}(M)}{2} \right) \!
                          \left(\ell-\frac32\right)\right]
 \notag\\ &\qquad
         + \dots,\\
 &S^+_\ell S^+_{\ell+1} =
  (-1)^{\ell+1} B_0^2  e^{i\sqfp\theta_+} \notag\\
 & \qquad\qquad
  + 2 B_0B_1  e^{i\sqfp\theta_+}
    \cos\!\left[\frac{\pi}{2}\!\left(\frac{1}{2}-|M|\right)\!\right]
\notag\\& \qquad\qquad\quad\times
 \cos\!\left[\sqp\phit_+
             + \pi\!\left(M+\frac{\mathrm{sgn}(M)}{2}\right)(\ell-1)\right]
  \notag\\
 &\qquad\qquad +\dots,
\end{align}
from which the correlation functions are calculated as
\begin{align}
 &\langle S_\ell^z S_{\ell'}^z \rangle =
  M^2 - \frac{K_+}{2\pi^2 |\ell'-\ell|^2} \notag\\
 &\qquad\qquad
  + B \frac{\cos[\pi(|M|-\frac12) (\ell'-\ell) ] }
           {|\ell'-\ell|^{K_+/2}} +\dots,\\
 &\langle S_\ell^+ S_{\ell+1}^+ S_{\ell'}^- S_{\ell'+1}^- \rangle = 
  B' \frac{ (-1)^{\ell'-\ell} }{ |\ell'-\ell|~^{2/K_+} } \notag\\ 
 &\qquad\qquad\qquad\qquad\quad
 - B'' \frac{\cos[\pi(|M|+\frac12)(\ell'-\ell) ] }
            { |\ell'-\ell|^{2/K_++K_+/2}} +\dots, 
\end{align}
with $B\propto A_1^2$, $B'\propto B_0^4$,
and $B''\propto B_0^2 B_1^2$. 
We note that the TLL phases
with similar power-law correlations,
called the nematic and SDW$_2$ phases, have also been discussed
for the model \eqref{eq:H} in a magnetic field,
for both ferromagnetic\cite{Hikihara08,Sato09,Sudan09,Heidrich09} 
and antiferromagnetic\cite{Okunishi03,Hikihara10_AF} $J_1$.
For small $|J_1|/J_2$ and $\Delta-1$, $K_+$ is close to unity, 
and the longitudinal (spin-density-wave; SDW) correlation 
decays more slowly than the nematic correlation. 
The TLL phase with a dominant SDW correlation and short-ranged
transverse spin correlation is called the SDW$_2$ state
in Refs.~\onlinecite{Hikihara08} and \onlinecite{Hikihara10_AF}. 
It is natural to assume that the partially polarized
phase at $\Delta>1$ in zero magnetic field 
is continuously connected to the SDW$_2$ phase
in a finite magnetic
field.\cite{Hikihara08,Sudan09,Heidrich09}  
With inter-chain couplings, the dominant quasi-long-range SDW correlation
is expected to evolve into a true long-range-order.\cite{Sato12} 
Since $K_+$ changes continuously in the TLL phases, 
it is also possible that the system crosses over to a region
with the dominant nematic correlation ($K_+>2$). 
It is known that such a region does appear
at high magnetic fields.\cite{Hikihara08,Sudan09,Heidrich09}

%%%%%%%%%%%%%%%%%%%%%%%%%%%%%%%%%%%%%%%%%%%%%%%%%
\section{Conclusions} \label{sec:conclusions}
%%%%%%%%%%%%%%%%%%%%%%%%%%%%%%%%%%%%%%%%%%%%%%%%%

In this paper, we have studied the ground-state properties of the
one-dimensional
spin-$\frac12$ frustrated ferromagnetic XXZ model \eqref{eq:H}. 
In the isotropic case $\Delta=1$, the nonmagnetic phase in the region
$-4<J_1/J_2<0$ was characterized as the Haldane dimer phase, 
in which the ground state has spontaneous ferromagnetic
dimerization and nonlocal string order.
We argued that the dimer order is associated with an emergent 
spin-$1$ degree of freedom on every other bond. 
In the easy-plane case $0\le \Delta<1$, 
the model displays a rich phase diagram as in Fig.~\ref{fig:phase}. 
Our previous works have revealed
the appearance of the gapless chiral
phase in a wide region for $-4<J_1/J_2<0$\cite{FSO10} 
and the unusual alternate appearance of the N\'eel and even-parity
dimer phases.\cite{FSF10} 
In this paper, we have newly discovered
narrow intermediate gapped phases 
in which the vector chiral order coexists with the dimer or N\'eel order. 
We described how the properties of the various phases can be captured
for $|J_1|/J_2\ll 1$ and general anisotropy $\Delta\ge 0$
by the Abelian bosonization formalism, as summarized in Table \ref{table:bos} 
(by continuity, the same qualitative description can be
extended to larger $|J_1|/J_2$). 

The Haldane dimer phase we found for $\Delta=1$ has only a very small
excitation gap and, with a weak easy-plane anisotropy, is easily
replaced by the gapless chiral phase.  With small inter-chain
couplings, the gapless chiral phase would evolve into a genuine spiral
long-range-order.  Therefore, the stable appearance of the gapless
chiral phase up to the close vicinity of the isotropic case $\Delta=1$
naturally explains why many quasi-one-dimensional cuprates with
ferromagnetic $J_1<0$ show the spiral magnetism and the associated
multiferroicity.\cite{FSO10} By contrast, it is also expected that the
small excitation gap ($\lesssim 0.06 J_2$; see
Sec.~\ref{subsubsec:dimer_corrlen}) in the Haldane dimer phase can be
enhanced by a coupling with phonons, due to the spin-Peierls mechanism
as is known in the antiferromagnetic $J_1$-$J_2$ chain compound
CuGeO$_3$.\cite{Hase93} It will be interesting to explore a
spin-Peierls transition to the Haldane dimer phase in quasi-1D
edge-sharing cuprates without a spiral magnetic order.  The present
study also raises the possibility of observing the chiral Haldane
dimer state, which shows no magnetic order but a spontaneous electric
polarization due to a vector chiral order of spins.

% In Rb$_2$Cu$_2$Mo$_3$O$_{12}$ (with the estimates $J_1=-138$ K and $J_2=51$ K), 
% no magnetic order has been observed down to 2 K.\cite{Hase04} 
% In LiCuSbO$_4$ (with the estimates $J_1=-75$ K and $J_2=34$ K), 
% a very recent experiment has confirmed the absence of a magnetic order down to 100 mK.\cite{Dutton12}

%[Memo]
% Rb$_2$Cu$_2$Mo$_3$O$_{12}$: J1=-138 K, J2=51 K, J_1/J_2=-2.7
% LiCuSbO$_4$: J1 = - 75 K, J2 = 34 K, J_1/J_2=-2.2

%%%%%%%%%%%%%%%%%%%%%%%%%%%%%%%%%%%%%%%%%%%%%%%%%
\acknowledgments
%%%%%%%%%%%%%%%%%%%%%%%%%%%%%%%%%%%%%%%%%%%%%%%%%

The authors thank S.\ Bhattacharjee, T.\ Hikihara, T.\ Momoi,
and K.\ Okunishi for stimulating discussions,   
and K.\ Nomura for his useful comment on the Lifshitz line. 
This work was supported by
Grants-in-Aid for Scientific Research (KAKENHI)
on Priority Areas ``Novel States of Matter induced by Frustration'' 
(No.\ 19052006, No.\ 20046016, No.\ 22014016) 
and on Innovation Areas ``Topological Quantum Phenomena'' (No.\ 22103005)  
and KAKENHI No.\ 21740295  
from MEXT of Japan, 
and KAKENHI No.\ 21740275, No.\ 24540338, and No.\ 24740253 
from Japan Society for the Promotion of Science. 
AF is grateful to the Galileo Galilei Institute for Theoretical
Physics and the Aspen Center for Physics for their hospitality,
where final edits of this paper were done.

% Furukawa: 22103005
% Sato: 22014016, 21740295
% Onoda: 19052006, 20046016, 21740275, 24740253
% Furusaki: 20046016, 24540338

%============================
% \begin{figure}[t]
% \begin{center}
% \includegraphics[width=0.48\textwidth]{.eps}
% \end{center}
% \caption{
% }
% \label{}
% \end{figure}
%============================

\appendix

%%%%%%%%%%%%%%%%%%%%%%%%%%%%%%%%%%%%%%%%%%%%%%%%%
\section{Derivation of the renormalization group equations \eqref{eq:RGeq}} \label{app:RG}
%%%%%%%%%%%%%%%%%%%%%%%%%%%%%%%%%%%%%%%%%%%%%%%%%

Here we briefly explain how the RG equations \eqref{eq:RGeq} are derived  
by using the perturbative RG method\cite{Cardy96}
and the operator product expansions (OPE) in the SU(2)$_1$
WZW theory. 

\newcommand{\zerov}{\bm{0}}

We first discuss the OPEs in the decoupled spin chains,
each described by the SU(2)$_1$ WZW theory. 
We drop the chain subscript $n=1,2$.  
The OPEs of the uniform spin components $M_{R/L}$ obey 
the well-known SU(2) current
algebra\cite{CFT96,Gogolin98,Shelton96,Starykh04,Starykh05} 
\begin{equation}
 M_{R/L}^a(x,\tau) M_{R/L}^b(\zerov) 
 = 
 \frac{ \delta^{ab} }{ 8\pi^2 z_{R/L}^2 }
 +
 \frac{ i\varepsilon^{abc} M_{R/L}^c(\zerov) }{ 2\pi z_{R/L} }
\end{equation}
with $z_{R/L}=v\tau\mp ix$. 
Here, $\varepsilon^{abc}$ is the fully antisymmetric tensor
with $\varepsilon^{123}=1$, 
and summation over repeated indices are assumed
throughout the appendix.
The OPEs present the singular terms that appear
when two operators at the points $(x,\tau)$ and
$\zerov =(0,0)$ are brought close together.

The OPEs of the uniform components $M^a_{R/L}$ with
the staggered components $N^a$ and the dimerization $\epsilon$
are given by\cite{Starykh05}
\begin{align}
 &M_{R/L}^a(x,\tau) N^b(\zerov) 
 = \frac{i}{4\pi z_{R/L}}
  \left[ \varepsilon^{abc} N^c(\zerov)
         \pm \delta^{ab}\epsilon(\zerov) \right],
 \label{eq:OPE of M with N}
 \\
 &M_{R/L}^a(x,\tau) \epsilon(\zerov)
 = \frac{ \mp iN^a(\zerov) }{ 4\pi z_{R/L} }.
 \label{eq:OPE of M with epsilon}
\end{align}
These equations imply that $M_{R/L}$ induce
mixing of $\Nvec$ and $\epsilon$.

Similar to Eqs.\ \eqref{eq:OPE of M with N} and
\eqref{eq:OPE of M with epsilon},
the OPEs among $\Nvec$ and $\epsilon$ can be
derived\cite{CFT96, Starykh05} by taking advantage of the well-known
spin-charge separation in 1D spin-$1/2$ Dirac fermions;
with bosonization, the charge and spin sectors
of Dirac fermions are described by
a free scalar boson and the SU(2)$_1$ WZW theory, respectively.
The use of fermionic fields simplifies the calculations of OPEs
in the WZW theory. 
For illustration, here we derive the OPE of two $\epsilon$'s.
We take the same conventions as used in the Appendix of
Ref.~\onlinecite{Starykh05}, 
and introduce the right- and left-moving fermionic fields $\Psi_{R/L,s}$
($s=\uparrow,\downarrow$),
which obey the OPEs
\begin{align}
 \Psi_{R/L,s} (x,\tau) \Psi_{R/L,s'}^\dagger (\zerov)
 = \frac{\delta_{ss'}}{2\pi z_{R/L}}. 
\end{align}
We define the fermionic staggered dimerization operator as 
\begin{align}
 \epsilon_F = \frac{i}{2} ( \Psi_{R s}^\dagger \Psi_{L s}
            - \Psi_{L s}^\dagger \Psi_{R s}). 
\end{align}
Using bosonization, one can show that $\epsilon_F$
is related to $\epsilon$ as
\begin{equation}
 \epsilon_F = \epsilon \cos (\sqtp \phi_\rho),
\end{equation}
where $\phi_\rho$ is the bosonic field of the charge sector.
We now assume that the charge sector is in the gapped Mott phase
where $\phi_\rho$ is locked ($\langle \phi_\rho \rangle=0$)  
as in the Hubbard chain at half-filling.
This allows us to identify $\epsilon_F$ with
$\lambda\epsilon$, where
$\lambda= \langle \cos (\sqtp \phi_\rho) \rangle$ is
a dimensionless constant of order unity.
The OPE of two $\epsilon$'s is then obtained from
the OPE of two $\epsilon_F$'s.

Performing all possible contractions of four fermion fields
(see Appendix A of Ref.~\onlinecite{Lin97}), 
the OPE of two $\epsilon_F$'s is calculated as  
\begin{equation}
\begin{split}
 \epsilon_F (x,\tau) \epsilon_F (\zerov) = {}&
 \frac14 \Psi_{Rs}^\dagger(x,\tau)\Psi_{Ls}(x,\tau) \Psi_{Ls'}^\dagger(\zerov)
 \Psi_{Rs'}(\zerov) \nonumber\\
 &{}+ (R\leftrightarrow L)\\
 ={}& \frac{1}{4\pi z_R z_L}
 + \frac{1}{4\pi} \left( \frac{\rho_R(\zerov)}{z_L}
 - \frac{\rho_L(\zerov)}{z_R} \right) \\
 & + \frac12 \Psi_{Rs}^\dagger(\zerov) \Psi_{Ls}(\zerov)
     \Psi_{Ls'}^\dagger(\zerov) \Psi_{Rs'}(\zerov)
\end{split}
\end{equation}
with $\rho_{R/L} = \Psi_{R/L,s}^\dagger \Psi_{R/L,s}$. 
The last term is related to the backscattering term: 
\begin{equation}
 \Psi_{Rs}^\dagger \Psi_{Ls} \Psi_{Ls'}^\dagger \Psi_{Rs'}
 = -2 \Mvec_R \cdot \Mvec_L - \frac12 \rho_R \rho_L,
\end{equation}
where the uniform components of the fermionic spin density are defined as
\begin{equation}
M^a_{R}=\frac12 \Psi_{Rs}^\dagger \sigma^a_{ss'} \Psi_{Rs'},
\qquad
M^a_{L}=\frac12 \Psi_{Ls}^\dagger \sigma^a_{ss'} \Psi_{Ls'}.
\end{equation}
After gapping out the charge sector, we can neglect the fluctuations
of $\rho_{R/L}$. 
Thus we obtain 
\begin{equation}
 \epsilon (x,\tau) \epsilon (\zerov) 
 = \frac{ 1 }{ 4\pi^2 \lambda^2 z_R z_L }
 - \frac1{\lambda^2} \Mvec_R(\zerov) \cdot \Mvec_L(\zerov).
\end{equation}
Similar calculations yield
\begin{align}
 N^a(x,\tau) N^b(\zerov) 
 ={}&\frac{ \delta^{ab} }{ 4\pi^2 \lambda^2 z_R z_L } \notag \\
 & + \frac{i \varepsilon^{abc}} {2\pi \lambda^2} 
   \left[ \frac{M_R^c(\zerov)}{z_L} + \frac{M_L^c(\zerov)}{z_R} \right]
 \notag \\
 & + \frac{1}{\lambda^2}\Ocal_{NN}^{ab} (\zerov), \label{eq:OPE_NN}
\end{align}
\begin{equation}
  N^a(x,\tau) \epsilon (\zerov) = 
 \frac{-i} {2\pi \lambda^2} 
 \left[ 
  \frac{M_R^a(\zerov)}{z_L} - \frac{M_L^a(\zerov)}{z_R}
 \right]
\end{equation}
% for the staggered components of the fermionic spin density operator,
% \begin{equation}
% N^a=\frac12 \sigma^a_{ss'}
% \left(\Psi^\dagger_{Rs}\Psi_{Ls'}+\Psi^\dagger_{Ls}\Psi_{Rs'}\right).
% \end{equation}
%
where $\Ocal_{NN}^{ab}$ in Eq.~\eqref{eq:OPE_NN} is expressed in terms of
fermionic fields as
\begin{equation}
 \Ocal_{NN}^{ab}=
\frac12 \sigma_{s_1s_2}^a \sigma_{s_3s_4}^b
\Psi_{Rs_1}^\dagger \Psi_{Ls_2} \Psi_{Ls_3}^\dagger \Psi_{Rs_4} . 
\end{equation}
For the current purpose, we only need the trace
(in the spin direction indices) of this term, 
which gives the backscattering term:
$ \Ocal_{NN}^{aa}=\Mvec_R \cdot \Mvec_L -\frac34 \rho_R \rho_L$. 

In the limit of weak interchain coupling $|J_1|\ll J_2$,
the OPEs of the perturbation operators in Eq.\ \eqref{eq:O_marginal} are
readily obtained from the OPEs of operators in each decoupled chain
described above.
Given the OPEs, one can write down the
corresponding one-loop RG equations.\cite{Cardy96} 
For example, if the OPE of marginal operators $\mathcal{O}_a$ and
$\mathcal{O}_b$ have the form
\begin{equation}
\mathcal{O}_a\mathcal{O}_b=
\frac{\lambda_{ab}^c}{(2\pi)^2 z_Rz_L}\mathcal{O}_c+\ldots,
\end{equation}
where $\lambda_{ab}^c$ are dimensionless constants,
then the one-loop RG equation for the perturbation $g_c\mathcal{O}_c$
has the contribution
\begin{equation}
\frac{dg_c}{dl}=-\frac{g_a g_b \lambda_{ab}^c }{4\pi v}+\ldots.
\end{equation}

%============================
\begin{figure}
\begin{center}
\includegraphics[width=0.48\textwidth]{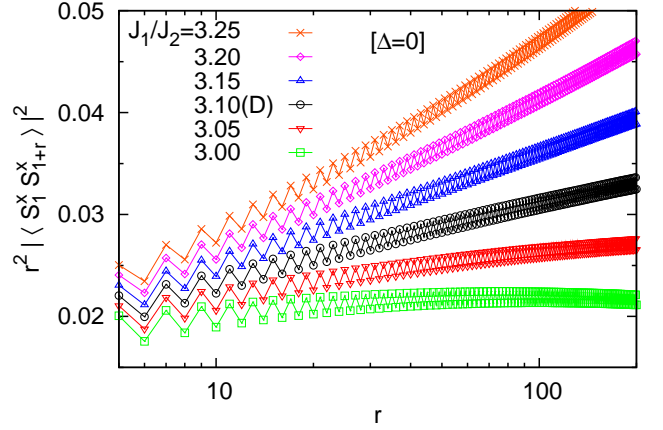}
\end{center}
\caption{(Color online)
Plots of $r^2|\langle S^x_1 S^x_{1+r}\rangle|^2$ 
for fixed $\Delta=0$ and various values of $J_1/J_2$ 
around the TLL-dimer transition point studied in Ref.~\onlinecite{Nomura94}. 
A logarithmic scale is used for the horizontal axis. 
The symbol ``D'' indicates the estimate of the transition point within the current analysis 
(with a precision of $0.05$), 
which agrees reasonably well with the previous accurate estimate\cite{Nomura94}  $J_2/J_1\approx 3.0893$. 
}
\label{fig:Cpm_z0}
\end{figure}
%============================

%%%%%%%%%%%%%%%%%%%%%%%%%%%%%%%%%%%%%%%%%%%%%%%%%
\section{TLL-dimer transition} \label{app:TLL-dimer}
%%%%%%%%%%%%%%%%%%%%%%%%%%%%%%%%%%%%%%%%%%%%%%%%%

In Sec.~\ref{subsec:numerics_easy-plane}, we determined the BKT transition
points between gapless and gapped chiral phases 
by observing the logarithmic correction in the spin correlation function
(Figs.~\ref{fig:Cpm_jm2} and \ref{fig:Cpm_z0p8jm3}). 
Here we test the validity of the method with
a simpler example. 
We consider the antiferromagnetic XY model with $J_1,J_2>0$ and $\Delta=0$. 
For large $J_1/J_2(\gtrsim 3)$, the system is in a Tomonaga-Luttinger
liquid (TLL) phase, 
in which the transverse spin correlation function behaves as\cite{Giamarchi04}
\begin{equation}
 \langle S_\ell^x S_{\ell'}^x \rangle 
 =  \frac{ A_0^x (-1)^{\ell'-\ell} }{ |\ell'-\ell|^\eta }
 -  \frac{A_1^x}{ |\ell'-\ell|^{\eta+1/\eta} } +\dots.
\end{equation}
Here $A_0^x$ and $A_1^x$ are non-universal constants.  The decay
exponent $\eta$ gradually increases as $J_1/J_2$ is decreased.
At $\eta=1$, a BKT transition from the TLL
to the singlet dimer phase occurs.
At the transition point, a multiplicative logarithmic
correction appears in the spin correlation
function:\cite{Giamarchi04,Kosterlitz74,Giamarchi88,Eggert96}
\begin{equation}
 \langle S_\ell^x S_{\ell'}^x \rangle 
 =  \frac{ A_0^x(-1)^{\ell'-\ell} }{ |\ell'-\ell| }
    \ln^{\frac12} (|\ell'-\ell|/a)  + \dots.
\end{equation}
In Fig.~\ref{fig:Cpm_z0}, we plot the function
$r^2|\langle S^x_1 S^x_{1+r}\rangle|^2$ 
for various $J_1/J_2$ around the BKT transition point. 
From the linear behavior as a function of $\ln r$,
we locate the BKT transition point. 
In this figure, the data points of
$J_1/J_2=3.10$ and $3.15$ exhibit almost
linear behavior.
It is not easy to decide which one of the two curves is
closer to the perfect linear dependence.
Here we choose the one with smaller correlations 
since the iTEBD method tends to underestimate correlations at large $r$. 
The determined point $J_1/J_2=3.10$ agrees reasonably well 
with the previous accurate estimate\cite{Nomura94} $J_2/J_1\approx 3.0893$.

%%%%%%%%%%%%%%%%%%%%%%%%%%%%%%%%%%%%%%%%%%%%%%%%%
%References
%%%%%%%%%%%%%%%%%%%%%%%%%%%%%%%%%%%%%%%%%%%%%%%%%

\newcommand{\etal}{{\it et al.}}
\newcommand{\PRL}[3]{Phys. Rev. Lett. {\bf #1}, \href{http://link.aps.org/abstract/PRL/v#1/e#2}{#2} (#3)}
\newcommand{\PRLp}[3]{Phys. Rev. Lett. {\bf #1}, \href{http://link.aps.org/abstract/PRL/v#1/p#2}{#2} (#3)}
\newcommand{\PRA}[3]{Phys. Rev. A {\bf #1}, \href{http://link.aps.org/abstract/PRA/v#1/e#2}{#2} (#3)}
\newcommand{\PRAp}[3]{Phys. Rev. A {\bf #1}, \href{http://link.aps.org/abstract/PRA/v#1/p#2}{#2} (#3)}
\newcommand{\PRB}[3]{Phys. Rev. B {\bf #1}, \href{http://link.aps.org/abstract/PRB/v#1/e#2}{#2} (#3)}
\newcommand{\PRBp}[3]{Phys. Rev. B {\bf #1}, \href{http://link.aps.org/abstract/PRB/v#1/p#2}{#2} (#3)}
\newcommand{\PRBR}[3]{Phys. Rev. B {\bf #1}, \href{http://link.aps.org/abstract/PRB/v#1/e#2}{#2} (R) (#3)}
\newcommand{\PRBRp}[3]{Phys. Rev. B {\bf #1}, \href{http://link.aps.org/abstract/PRB/v#1/p#2}{R#2} (#3)}
\newcommand{\arXiv}[1]{arXiv:\href{http://arxiv.org/abs/#1}{#1}}
\newcommand{\condmat}[1]{cond-mat/\href{http://arxiv.org/abs/cond-mat/#1}{#1}}
\newcommand{\JPSJ}[3]{J. Phys. Soc. Jpn. {\bf #1}, \href{http://jpsj.ipap.jp/link?JPSJ/#1/#2/}{#2} (#3)}
\newcommand{\JPSJS}[3]{J. Phys. Soc. Jpn. Suppl. {\bf #1}, \href{http://jpsj.ipap.jp/link?JPSJS/#1S/#2/}{#2} (#3)}
\newcommand{\PTPS}[3]{Prog. Theor. Phys. Suppl. {\bf #1}, \href{http://ptp.ipap.jp/link?PTPS/#1/#2/}{#2} (#3)}
\newcommand{\hreflink}[1]{\href{#1}{#1}}


\begin{thebibliography}{99}
\bibitem{Comment_dimer}
 This phase was called the triplet dimer phase
 in Refs.~\onlinecite{FSO10,SFOF11}.
\bibitem{note:iTEBD}
For example, in the singlet dimer phase with two-fold degenerate
ground state, the variational state converges to one of the two
symmetry-broken ground states, not to an arbitrary linear combination
of the two.
\bibitem{Comment_O1_bos} The negative sign in the first line of
 Eq.~\eqref{eq:O1_bos} was often missed in previous studies.  This
 sign is crucial in determining the field locking positions in
 different dimer phases as in Eqs.~\eqref{eq:lock_dimer} and
 \eqref{eq:lock_tridimer}.  To determine this sign correctly, it is
 crucial to notice that $e^{i\sqtp\theta_n}$ and $\cos(\sqtp\phi_n)$
 in Eq.~\eqref{eq:Sp_bos} anticommute with each other.
This is derived from
\begin{align}
 e^{i\sqtp\theta_n(x)} e^{\pm i\sqtp\phi_n(x)} = {}&
 e^{\pm i\sqtp\phi_n(x)} e^{i\sqtp\theta_n(x)} \nonumber\\
 & \times e^{\mp 2\pi [\theta_n(x),\phi_n(x)]} \nonumber \\
 = & - e^{\pm i\sqtp\phi_n(x)} e^{i\sqtp\theta_n(x)}, \nonumber
\end{align}
 where Eq.~\eqref{eq:comm_bos} and $Y(0)=1/2$
 are used.
\bibitem{Comment_White96} In Ref.~\onlinecite{White96}, the
 correlation length was defined in unit of two lattice spacings, and
 should be doubled when comparing with the current data.
\bibitem{Comment_ES}
 Because of the $SU(2)$ symmetry of the model, we can further confirm
 that all the entanglement levels are even- or odd-fold degenerate
 in left and right panels of Fig.~\ref{fig:ES}, respectively.
\end{thebibliography}
\end{document}